\documentclass[12pt]{article}

\usepackage{amsmath,amsfonts,enumerate,array}
\usepackage{graphicx}
\usepackage[usenames]{color}
\usepackage[english]{babel}
\usepackage{fancybox}
\usepackage{chngpage} 
\usepackage{subcaption,cite,xcolor,lipsum}

\renewcommand{\a}{\alpha}	
\newcommand{\g}{\gamma}		

\newcommand{\s}{\sigma}		

\newcommand{\e}{\epsilon}

\newcommand{\m}{\mu}

\newcommand{\nn}{\nonumber\\} 		
\newcommand{\h}{{1\over2}}				

\def\h{\mathbb{h}}

\def\one{{\hbox{\kern+.5mm 1\kern-.8mm l}}}
\def\zero{{\hbox{0\kern-1.5mm 0}}}

\renewcommand{\ell}{l}

\newcommand{\pa}{\partial}
\newcommand{\ba}{\begin{align}}
\newcommand{\ea}{\begin{align}}

\newcommand\blfootnote[1]{%
  \begingroup
  \renewcommand\thefootnote{}\footnote{#1}%
  \addtocounter{footnote}{-1}%
  \endgroup
}

\newcommand{\captionfonts}{\small}

\makeatletter  
\long\def\@makecaption#1#2{%
  \vskip\abovecaptionskip
  \sbox\@tempboxa{{\captionfonts #1: #2}}%
 \ifdim \wd\@tempboxa >\hsize
    {\captionfonts #1: #2\par}
  \else
    \hbox to\hsize{\hfil\box\@tempboxa\hfil}%
  \fi
  \vskip\belowcaptionskip}
\makeatother   

\usepackage{mciteplus}

\usepackage[colorlinks=true,      linkcolor=blue,      urlcolor=blue,      filecolor=blue,      citecolor=blue,
      pdfstartview=FitH,     pdfpagemode=UseNone,      bookmarksopen=true]{hyperref}  
      
\usepackage[all]{hypcap}  
      
\title{Full action of two deformation operators in the D1D5 CFT}

\begin{document}

\numberwithin{equation}{section}

\def\b{\bigskip}
\def\p{\partial}
\def\h{{1\over 2}}
\def\be{\begin{equation}}
\def\bea{\begin{eqnarray}}
\def\ee{\end{equation}}
\def\eea{\end{eqnarray}}
\def\nn{\nonumber \\}
\def\m{\medskip}
\def\r{\rightarrow}
\def\t{\tilde}

\vspace{16mm}

 \begin{center}
{\LARGE Lifting of D1-D5-P states}

\vspace{18mm}
{\bf   Shaun Hampton$^1$\blfootnote{$^{1}$hampton.197@osu.edu},  Samir D. Mathur$^2$\blfootnote{$^{2}$mathur.16@osu.edu} and   Ida G. Zadeh$^3$\blfootnote{$^{3}$zadeh@math.ethz.ch}
\\}
\vspace{15mm}
${}^{1,2}$Department of Physics,\\ The Ohio State University,\\ Columbus,
OH 43210, USA\\ \vspace{8mm}
${}^{3}$Department of Mathematics, \\ ETH Zurich, \\ CH-8092 Zurich, Switzerland\\

\vspace{8mm}
\end{center}

\vspace{4mm}

\thispagestyle{empty}
\begin{abstract}

\vspace{3mm}

We consider states of the D1-D5 CFT where only the left-moving sector is excited. As we deform away from the orbifold point, some of these states will remain BPS while others can `lift'. We compute this lifting for a particular family of D1-D5-P states, at second order in the deformation off the orbifold point. We note that the maximally twisted sector of the CFT is special: the covering surface appearing in the correlator can only be genus one while for other sectors there is always a genus zero contribution.  We use the results to argue that fuzzball configurations should be studied for the full class including both extremal and near-extremal states; many extremal configurations may be best seen as special limits of near extremal configurations.

\end{abstract}
\newpage

{\hypersetup{linkcolor=black}
\tableofcontents}

\section{Introduction}\label{Intro}

One of the most useful examples of a black hole is the hole made with D1, D5, and P charges in string theory. The microscopic entropy for these charges, $S_{micro}$,  agrees with the Bekenstein entropy, $S_{bek}$, obtained from the classical gravity solution with the same charges \cite{sv,cm}. Further, a weak coupling computation of radiation from the branes, $\Gamma_{micro}$,  agrees with the Hawking radiation from the gravitational solution, $\Gamma_{hawking}$ \cite{dmcompare,maldastrom}. 

AdS$_3$/CFT$_2$ duality  \cite{maldacena, gkp, witten} states that the near horizon dynamics of the black hole is described by a 1+1 dimensional CFT called the D1-D5 CFT. The momentum charge P is carried by left moving excitations of this CFT. The CFT has a `free point' called the `orbifold CFT', where the theory can be described using free bosons and free fermions on a set of twisted sectors \cite{Vafa:1995bm,Dijkgraaf:1998gf,orbifold2,Larsen:1999uk,Arutyunov:1997gt,Arutyunov:1997gi,Jevicki:1998bm}, see \cite{David:2002wn} for a review of the D1-D5 brane system. 

 At the orbifold point all states which have only left moving excitations are BPS; i.e. they have energy equal to their charge. This need not be true as we deform the theory along some direction in the moduli space of  the D1-D5 CFT. Some of the states which were BPS at the orbifold point will remain BPS, while others can pair up and `lift'.

In this paper we will look at a specific family of D1-D5-P states which are BPS at the orbifold point but which lift as we move away from this free point towards the supergravity description of the black hole. We use conformal perturbation theory to compute the lifting at quadratic order in the coupling $\lambda$. The form of this lifting will tell us about the behavior of string states in the gravity dual, and shed light on the nature of the fuzzball configurations that describe black hole microstates \cite{fuzzballs_i,fuzzballs_ii,fuzzballs_iii,fuzzballs_iv,fuzzballs_v}. 

We now summarize the set-up and the main results.

\subsection{The D1-D5 CFT}

We consider type IIB string theory compactified as
\be
M_{9,1}\r M_{4,1}\times S^1\times T^4\ .
\ee
We wrap $n_1$ D1 branes on $S^1$ and $n_5$ D5 branes on $S^1\times T^4$. The bound states of these branes generate the D1-D5 CFT,  which is a 1+1 dimensional field theory living on the cylinder made from  the $S^1$ and time directions. This theory is believed to have an orbifold point, where we have
\be
N=n_1n_5
\ee
copies of a $c=6$ free CFT. The free CFT is made of $4$ free bosons and $4$ free fermions in the left-moving sector and likewise in the right-moving sector. The free fields are subject to an orbifold symmetry generated by the  group of permutations $S_N$; this leads to various twisted sectors around the circle $S^1$. The field theory is a CFT with small ${\cal N}=4$ supersymmetry in each of the left and right-moving sectors; thus the left sector has chiral algebra generators $L_n, G^{\pm}_{\dot A,r}, J^a_n$ associated with the stress-energy tensor, the supercurrents, and the $\mathfrak{su}(2)$ R-currents (the right-moving sector has analogous generators). The small $\mathcal N=4$ superconformal algebra and our notations are outlined in appendix \ref{app_cft}.

\subsection{The states of interest}\label{introsec}

\begin{figure}
\begin{center}
\includegraphics[width=71mm]{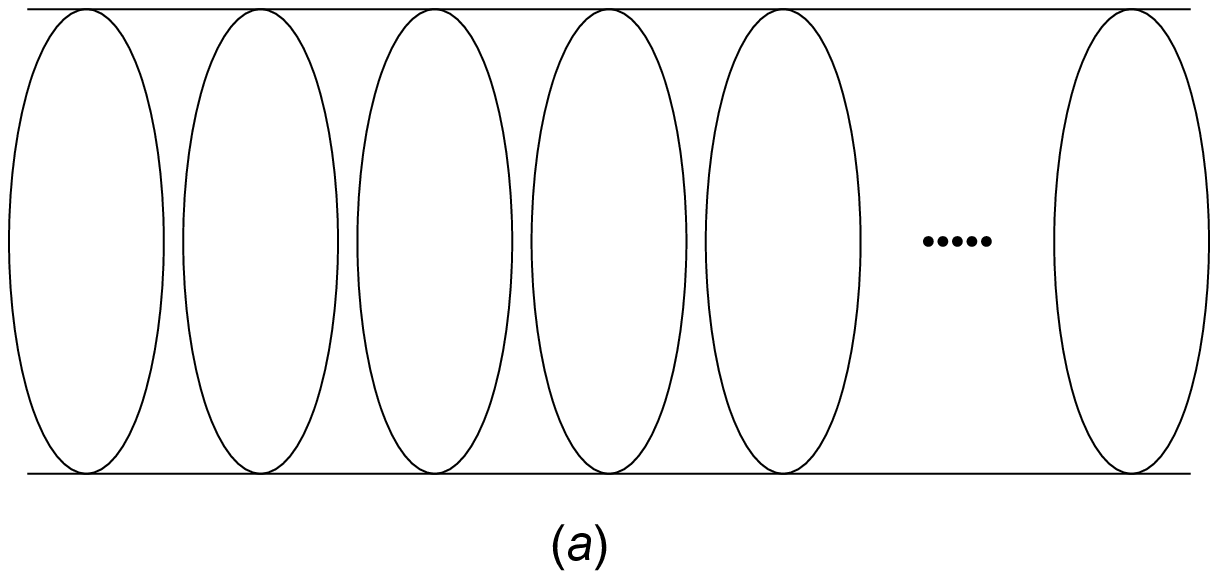}$\qquad\qquad$
\includegraphics[width=71mm]{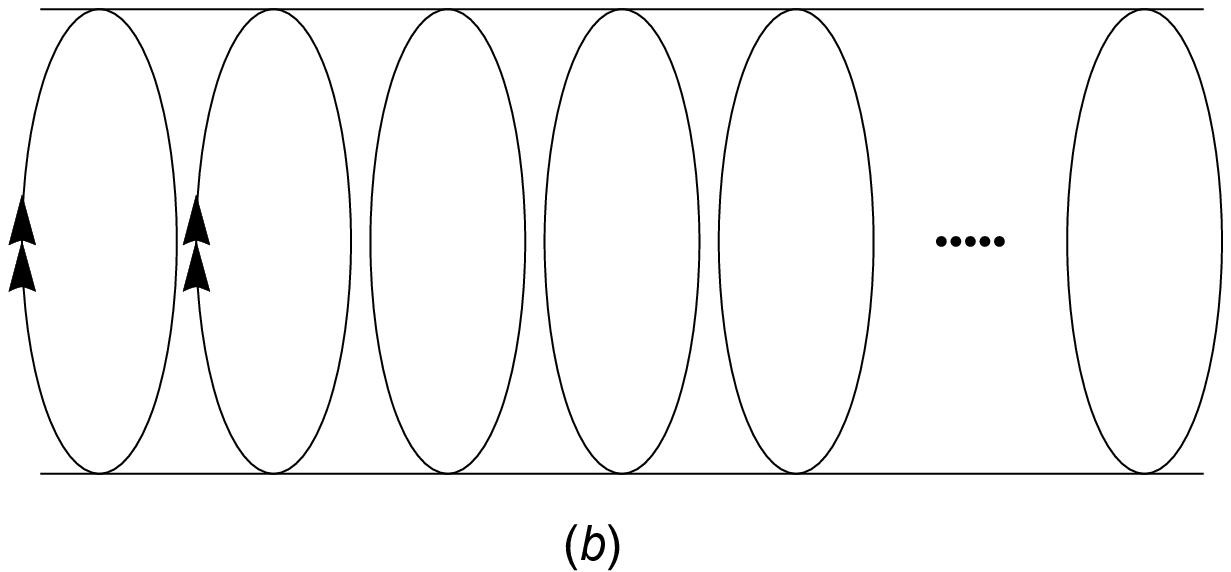}
\end{center}
\label{singlwinding}
\caption{(a)  $N$ singly-wound component strings in their vacuum state wrapping the $S^1$; this gives the vacuum of the theory.  (b) $n$ of these $N$ copies have been excited by the application of current operators.}
\label{fig_singlwindingexcited}
\end{figure}

Consider the untwisted sector, i.e. the sector where each copy of the $c=6$ CFT is singly wound around the $S^1$.  The $N$ copies of the $c=6$ CFT can be depicted  by $N$ separate circles; we sketch this in fig. \ref{fig_singlwindingexcited}. We consider the NS sector. If all the copies are unexcited, we get the vacuum state $|0\rangle$ as shown in the left panel of the figure; the gravity dual of this state is $AdS_3 \times S^3\times T^4$.

We will now consider a set of excited states, proceeding in the following steps:

\b

(i) Consider the state where $N-1$ copies are in the NS vacuum state, and one of the copies is excited by application of the operator
\be
 J^{+}_{-(2m-1)}\dots J^{+}_{-3}J^{+}_{-1}
\label{threeq}
\ee
on the NS vacuum. This is illustrated in the right panel of fig. \ref{fig_singlwindingexcited}. At the orbifold point this operator has quantum numbers
\be
(h,\bar h)=(m^2,0)\ ,\qquad(j,\bar j)=(m,0)\ .
\ee
Thus the energy of the state at the orbifold point  is
\be
E_{\mathrm{orbifold}}=h+\bar h = m^2\ .
\label{four}
\ee
At the orbifold point, the excitation is BPS since the right movers of the CFT are  in the supersymmetric ground state on all copies. 

We will argue that when we move to the supergravity domain, this state can be heuristically described by  a string localized at the center of the AdS space. Because the excitation is a string rather than a supergravity quantum, the energy will change away from the orbifold point (\ref{four}); we write the extra  energy as
\be
E-E_{\mathrm{orbifold}}=\Delta E\ .
\ee

\b

(ii) Suppose we place the excitation (\ref{threeq}) on {\it two} copies of the CFT. At the orbifold point we have
\be
(h,\bar h)=(2m^2,0)\ ,\qquad(j,\bar j)=(2m,0)\ ,
\ee
and
\be
E_{\mathrm{orbifold}}=2m^2\ .
\ee
At the supergravity point in the dual theory, our heuristic picture will  have {\it two} strings placed at the center of AdS. Each string will have an extra energy $\Delta E$ as before. But there will also be some gravitational attraction between these strings, which will lower the energy by some amount $\Delta E_{\mathrm{grav}}$. This suggests that the energy at the supergravity point will have the schematic form
\be
E-E_{\mathrm{orbifold}}=2\Delta E-E_{\mathrm{grav}}\ .
\ee

\b

(iii) Now suppose we place the excitation (\ref{threeq}) on $n$ out the the $N$ copies. In the dual gravity description we have $n$ strings. We get a positive energy from each of the $n$ strings, and a negative contribution to the energy from the attraction between each pair of strings. Thus the total energy has the schematic form
\be
E_n-E_{\mathrm{orbifold},n }=n\Delta E - {n(n-1)\over 2} E_{\mathrm{grav}}\ ,
\label{five}
\ee
where we have added a subscript  $n$ to the energies to indicate the number of copies which have been excited.

\b

(iv) Finally we place the excitation (\ref{threeq}) on {\it all} the $N$ copies of the $c=6$ CFT. In this situation we  know the energy $E$ exactly, because this state is obtained by a spectral flow of the vacuum by $2m$ units. We have
\be
E_N-E_{\mathrm{orbifold},N}=0
\ee

\b

\begin{figure}
\begin{center}
\includegraphics[width=80mm]{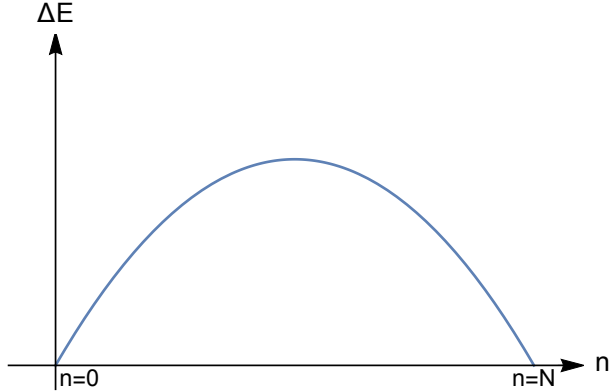}
\end{center}
\caption{A schematic plot showing the energy lift when $n$ out of the $N$ copies of the $c=6$ CFT  are excited.  The lift vanishes when $n=0$ and when $n=N$: the $n=0$ state is the vacuum and the $n=N$ state is a spectral flow of the vacuum.}
\label{fig_deltaE}
\end{figure}

To summarize, the quantity  $E_n-E_{\mathrm{orbifold},n}$ should have the schematic behavior depicted in fig. \ref{fig_deltaE}: it vanishes at $n=0$, rises for low values of $n$, then falls back to zero at $n=N$. 

\subsection{The results}

In this paper we study the deformation of the D1-D5 CFT off the orbifold point towards the supergravity point, upto second order in the coupling $\lambda$. We show that, at this order, the energy $E_n$ of the family of the states we consider in eq. (\ref{threeq}) indeed has the form depicted in fig.\ref{fig_deltaE}:
\be
\langle ( E_n-E_{\mathrm{orbifold}}) \rangle =\lambda^2 \frac{\pi^{3\over 2}}{2}\,{\Gamma[m^2-{1\over2}]\over \Gamma[m^2-1]}\,n(N-n)\ ,
\label{wwone}
\ee
where $m$ is the number of the R-currents in the initial state (\ref{threeq}), see section \ref{section7} for details.

We also consider the case where the $N$ copies of the CFT are grouped into twist sectors with winding $k$ each, and then excited in a manner similar to that discussed above, We again find an expression of the energy lift of the form (\ref{wwone}). 

Finally, we note a general property of the computation of lifting at second order. If the deformation operators join two component strings and then break them apart, the covering surface arising in the computation has genus 0. If on the other hand  the deformation operators break and then rejoin a component string,  the covering space arising in the computation has genus 1. The maximally twisted sector can only exhibit the second possibility; this suggests that the large class of unlifted states needed to explain black hole entropy may lie in this sector.

There are several earlier works that have studied conformal perturbation theory, the lifting of the states, the acquiring of anomalous dimensions, and the issue of operator mixing,  in particular in the context of the D1-D5 CFT;  see for example \cite{Pakman:2009mi,Avery:2010er,Burrington:2012yq,Burrington:2014yia,Gaberdiel:2015uca,Burrington:2017jhh,Carson:2016uwf}.

\subsection{The plan of the paper}
In section \ref{section2} we outline the computation that gives the lift to second order; in particular we explain why the issue of operator renormalization does not arise to this order in our problem.

 In section \ref{section3} we describe the deformation operator and the states whose lift we are interested in.

In section \ref{section4} we compute  the vacuum correlation function of two twist-2 operators; we call this the `base' amplitude, as it appears as a starting element in the computation of all other correlation functions.

In section \ref{section5} we compute the lifting of energies of the states under consideration.

In section \ref{section6} we consider global modes and use our approach to show that they are not lifted under conformal perturbation, to the order we study, as expected.

In section \ref{section7} we perform the needed combinatorics to extend our result to the case where we have an arbitrary number $N$ of component strings.

In section \ref{section8} we compute the lift for the case where the component strings on the initial state are grouped into sets with twist $k$ each.

In section \ref{section9} we analyze the nature of the covering space  in different instances of second order perturbation theory, and find a special role for the maximally twisted sector.

Section  \ref{section10} is a general discussion, where we state the physical implications of our results.

\section{Outline of the method}\label{section2}
In this section we first outline the conformal perturbation theory approach that we will use to compute the lifting of energies. We then derive a general expression for lifting to second order.

\subsection{Conformal perturbation theory on the cylinder}\label{outline}
We proceed in the following steps:

\b

(a) Suppose we have a conformal field $\phi$ with left and right-moving dimensions $(h, 0)$. On the plane, the 2-point function is
\be
\langle \phi(z) \phi(0) \rangle_0 ={1\over z^h}\ ,
\ee
where the subscript $``0"$ corresponds to the unperturbed theory. After a perturbation of the CFT, the conformal dimensions can change to $(h+\delta h, \delta h)$. The left and right dimensions must increase by the same amount, since $h-\bar h$ must always be an integer for the operator to be local. The operator $\t\phi$ in the perturbed theory, having a well defined dimension, will also in general be different from $\phi$. So we should write
\be\label{phiphi_i}
\phi=\t\phi+\delta \t\phi\ .
\ee
We will see, however, that to the order where we will be working, the correction $\delta \t\phi$ will not be relevant (see eq.s (\ref{phiphi}), (\ref{bone_i}), (\ref{bone}), and footnote \ref{fn_bone}). We can then write
\be
\langle \phi(z) \phi(0) \rangle_{\mathrm{pert}} ={1\over z^h}{1\over |z|^{2\delta h}}= {1\over z^h}e^{-2\delta h\log |z|}\approx {1\over z^h} ( 1-2\delta h\log |z|)\ ,
\ee
where the subscript ``pert" corresponds to the perturbed theory. Thus the perturbation to the 2-point function has a correction term of the form $ \sim z^{-h}\log|z|$, and the perturbation to the  dimension $\delta h$ can be read off from the coefficient of this term. For more details on the analyses of conformal perturbation theory in two and higher dimensional CFTs see, e.g. \cite{kadanoff,Dijkgraaf:1987jt,Cardy:1987vr,Eberle:2001jq,Gaberdiel:2008fn,Berenstein:2014cia,Berenstein:2016avf}.

\b

(b) We will find it convenient to work on the cylinder rather than the plane, so let us see how the expressions in (a) change when  we work on the cylinder.  The cylinder coordinate is given by
\be
z=e^w, ~~~w=\tau+i\sigma\ .
\ee
Consider the  state $|\phi\rangle$ corresponding to the operator $\phi$; we assume that this state is normalized as
\be\label{unit_norm}
\langle \phi |\phi\rangle_0 =1\ .
\ee
As will become clear below, we can ignore the change $\delta \t\phi$ in this state itself. Let us also assume for the moment  that the energy level of $\phi$ is nondegenerate. We  place the state $|\phi\rangle$ at $\tau=-{T\over 2}$ and the conjugate state $\langle \phi |$ at $\tau={T\over 2}$. We compute the amplitude $A(T)$ for transition between these two states. In the unperturbed theory, we have for our operator $\phi$ with dimensions $(h,0)$:
\be\label{amp_unpert}
A(T)=\langle \phi |e^{-H^{(0)} T}|\phi\rangle = e^{-hT}\ ,
\ee
where the energy of the state is $E=h+\bar h = h$, and $H^{(0)}$ is the Hamiltonian in the unperturbed theory. After the perturbation, we get
\be\label{amp_pert}
A(T)+\delta A(T) = \langle \phi |e^{-(H^{(0)}+\delta H) T}|\phi\rangle= e^{-(h+2\delta h)T}\approx e^{-hT}-2 \delta h\,  T e^{-hT}\ .
\ee
Thus, we can read off $\delta h$ from the coefficient of $Te^{-hT}$ in $\delta A(T)$. 

\b

(c) With these preliminaries, we now set up the formalism for situation that we actually have.  In our problem, the space of operators with dimension $(h, 0)$ is degenerate. Thus, we consider the case where we have operators $\phi_a, a=1, \dots,n$, all with the same dimension $(h,0)$. Let the remaining operators having well defined scaling dimensions be  called $\phi_\mu$; there will in general be an infinite number of the $\phi_\mu(\lambda)$, with dimensions going all the way to infinity. These operators are normalized as
\be
\langle \phi_{a}|\phi_{b}\rangle=\delta_{ab}\ ,\qquad\langle \phi_{\mu}|\phi_{\nu}\rangle=\delta_{\mu\nu}\ ,\qquad\langle \phi_{a}|\phi_{\mu}\rangle=0\ .
\label{none}
\ee
After the perturbations, there will be a different set of operators which have well defined scaling dimensions; let these operators be denoted by a tilde on top. We separate these operators into two classes. The first class is the operators that are deformations of the degenerate set $\phi_a, a=1, \dots n$. We call these deformed operators $\t \phi_{a'}(\lambda), a'=1, \dots n$, where we have explicitly noted the dependence of these operators on the coupling $\lambda$.  The second class is comprised of the remaining operators in the deformed theory which have well defined scaling dimensions; let us call these $\t\phi_\mu(\lambda)$.  We assume that these operators are normalized
\be
\langle \t\phi_{a'}(\lambda)|\t\phi_{b'}(\lambda)\rangle=\delta_{a'b'}\ ,\qquad\langle \t\phi_{\mu}(\lambda)|\t\phi_{\nu}(\lambda)\rangle=\delta_{\mu\nu}\ ,\qquad\langle \t\phi_{a'}(\lambda)|\t\phi_{\mu}(\lambda)\rangle=0\ .
\ee
The $\t\phi_{a'}$ have conformal dimensions of $(h+\delta h_{a'}(\lambda), \delta h_{a'}(\lambda))$.  The energies of the unperturbed states $\phi_{a'}$ and the perturbed states $\t\phi_{a'}$ are therefore
\be
E=h+\bar h=h\ , ~~~\t E_{a'}= h + 2\delta h_{a'}(\lambda)\ ,
\ee 
We expand the perturbed energies as  
\be
\t E_{a'}(\lambda)=E+\lambda E^{(1)}_{a'}+\lambda^2 E^{(2)}_{a'}+\cdots\ ,
\ee
\be
\t E_{\mu'}=E_{\mu'}+\lambda E^{(1)}_{\mu'}+\lambda^2 E^{(2)}_{\mu'}+\cdots \ .
\ee

Let us now consider the expansions of operators themselves. We can write
\bea
\t\phi_{a'}(\lambda)&=&\t C_{a'a}(\lambda)\phi_a+\t D_{a'\mu}(\lambda)\phi_\mu\ ,\nn
\t\phi_{\mu'}(\lambda)&=&\t F_{\mu'a}(\lambda)\phi_a+\t G_{\mu'\nu}(\lambda)\phi_\nu\ ,
\eea
where $\t C_{a'a}$, $\t D_{a'\mu}$, $\t F_{\mu'a}$, and $\t G_{\mu'\nu}$ are  $\lambda$-dependent expansion coefficients. We can invert these expansions to write
\bea
\phi_{a}&=&C_{aa'}(\lambda)\t \phi_{a'}(\lambda)+D_{a\mu'}(\lambda)\t \phi_{\mu'}(\lambda)\ ,\nn
\phi_{\mu}&=&F_{\mu a'}(\lambda)\t \phi_{a'}(\lambda)+G_{\mu\nu'}(\lambda)\t \phi_{\nu'}(\lambda)\ .
\eea
Finally, we expand the coefficients above in powers of $\lambda$: 
\bea
C_{aa'}(\lambda)&=&C^{(0)}_{aa'}+\lambda C^{(1)}_{aa'}+\lambda^2 C^{(2)}_{aa'}+\dots\nn
D_{a\mu'}(\lambda)&=&D^{(0)}_{a\mu'}+\lambda D^{(1)}_{a\mu'}+\lambda^2 D^{(2)}_{a\mu'}+\dots
\eea
Thus, in particular $\phi_a$ can be expanded as
\be\label{phiphi}
\phi_a=C^{(0)}_{aa'}\t\phi_{a'}+\lambda C^{(1)}_{aa'} \t\phi_{a'}+\lambda^2 C^{(2)}_{aa'}  \t \phi_{a'}+\cdots +\lambda D^{(1)}_{a\mu} \t \phi_\mu +\cdots\ .
\ee
The condition (\ref{none}) gives at leading order
\be
C^{(0)}_{aa'} C^{(0)*}_{ba'}=\delta_{ab}\ .
\label{nfive}
\ee 

The reason all these preliminaries are needed is that when computing an amplitude in  perturbation theory we find ourselves in the following situation. The operators in the amplitude  are taken to be the unperturbed operators $\phi_a, \phi_\mu$, since these are the ones with well understood and explicit constructions. But the operators that have well defined scaling dimensions are the $\t\phi_{a'}, \t\phi_{\mu'}$, which are not explicitly known. Thus we would compute an amplitude of the type
\be
A_{ab}(T)\equiv\Big\langle\phi_b({\tfrac T 2})\Big|e^{-(H^{(0)}+\delta H(\lambda)) T}\Big| \phi_a(-{\tfrac T 2})\Big\rangle\ .
\label{ccone}
\ee
Here the operators $\phi_a, \phi_b$ are operators in the unperturbed theory, and therefore explicitly known to us. But these unperturbed operators do not give eigenstates of the full Hamiltonian  $H^{(0)}+\delta H(\lambda)$; the latter eigenstates correspond to the perturbed operators $\t\phi_{a'}, \t\phi_{\mu'}$. Thus we have
\bea
\langle\t\phi_{b'}({\tfrac T 2})\Big|e^{-(H^{(0)}+\delta H(\lambda)) T}\Big| \t\phi_{a'}(-{\tfrac T 2})\Big\rangle&=&e^{-\t E_{a'} T}\delta_{a'b'}\ ,\nn
\langle\t\phi_{\nu'}({\tfrac T 2})\Big|e^{-(H^{(0)}+\delta H(\lambda)) T}\Big| \t\phi_{\mu'}(-{\tfrac T 2})\Big\rangle&=&e^{-\t E_{\mu'} T}\delta_{\mu'\nu'}\ ,\nn
\langle\t\phi_{a'}({\tfrac T 2})\Big|e^{-(H^{(0)}+\delta H(\lambda)) T}\Big| \t\phi_{\mu'}(-{\tfrac T 2})\Big\rangle&=&0\ .
\eea

Substituting the expansions (\ref{phiphi}) in eq. (\ref{ccone}), we find
\bea\label{AabT}
&&A_{ab}(T)\equiv\Big\langle\phi_b({\tfrac T 2})\Big|e^{-(H^{(0)}+\delta H(\lambda)) T}\Big| \phi_a(-{\tfrac T 2})\Big\rangle \nn
&&\!\!=\Big(C^{(0)*}_{ba'}+\lambda C^{(1)*}_{ba'}+\lambda^2 C^{(2)*}_{ba'} +\dots\Big)\Big(C^{(0)}_{aa'}+\lambda C^{(1)}_{aa'}+\lambda^2 C^{(2)}_{aa'} +\dots\Big)e^{-(E+\lambda E^{(1)}_{a'}+\lambda^2 E^{(2)}_{a'}+\dots )T}\nn
&&\!\!+\,\lambda^2 D^{(1)*}_{b\mu}D^{(1)}_{a\mu}e^{-(E_\mu+\lambda E^{(1)}_{\mu}+\lambda^2 E^{(2)}_{\mu}+\dots )T} +\cdots\ .
\eea
In general amplitudes like $A_{ab}(T)$ are functions of the fields like $\phi_a, \phi_b$ placed at the upper and lower time slices, the time interval $T$ between the slices, and the coupling $\lambda$. From the set of such amplitudes, we can extract the perturbed dimensions of the theory.  We will do this below, but first we note that it is convenient  to expand the above amplitude  in powers of $\lambda$ 
\be\label{Aab_define}
A_{ab}(T)=A^{(0)}_{ab}+\lambda A^{(1)}_{ab} +\lambda^2 A^{(2)}_{ab}+\cdots\ .
\ee

\b

(d) We first look at the coefficient of $\lambda Te^{-ET}$ in (\ref{AabT}). This coefficient is found to be
\be
-C^{(0)}_{aa'}E^{(1)}_{a'} C^{(0)*}_{ba'}\ .
\ee
We can write the above relation in matrix form, defining $(\hat A^{(1)})_{ab}=A^{(1)}_{ab}$, $(\hat C^{(0)})_{aa'}=C^{(0)}_{aa'}$, and $(\hat{E}^{(1)})_{a'b'}=\delta_{a'b'}  E^{(1)}_{a'}$. This gives
\be
 \hat A^{(1)}  \r - T e^{-ET} \hat C^{(0)} \hat{E}^{(1)} \hat C^{(0)\dagger}\ .
\label{bone_i}
\ee
 where the arrow indicates that we are writing only the coefficient of $Te^{-ET}$ in $ \hat A^{(1)} $.

We now note that, in our problem, the amplitude $A_{ab}(T)$ has no terms at $O(\lambda)$. This is because, as we will see in the next subsection, the deformation operator $D$ which perturbs the theory away from the orbifold point is in the twist 2 sector, while the states $|\phi_a\rangle$ and $|\phi_b\rangle$ are in the untwisted sector. The 3-point function $\langle\phi_b|D|\phi_a\rangle$ then vanishes due to the orbifold group selection rules. From eq. (\ref{nfive}) we see that $\hat C$ is unitary. Thus the vanishing of the above contribution tells us that $\hat{E}^{(1)}=0$; i.e. $E^{(1)}_{a'}=0$ for all $a'\in\{1, \dots,n\}$.

Now we look at  the coefficient of $\lambda^2 Te^{-ET}$ in $ A_{ab}(T)$ in eq. (\ref{AabT}). We find 
\be\label{A2abT}
 A^{(2)}_{ab} \r   - Te^{-ET} \left ( C^{(0)}_{aa'}C^{(0)*}_{ba'} E^{(2)}_{a'}\right )\ .
\ee
In matrix form, this reads 
\be
 \hat A^{(2)}  \r - T e^{-ET} \hat C^{(0)} \hat { E^{(2)}} \hat C^{(0)\dagger}\ .
\label{bone}
\ee
Thus, if we compute the matrix $\hat A^{(2)}$ and look at the coefficient of $-Te^{-ET}$,  then the eigenvalues of this matrix give the corrections to the energies upto $O(\lambda^2)$ :
\be\label{E2T}
 E^{(2)}_{a'}=2\delta h_{a'}\ ,
\ee
 and the eigenvectors give the linear combinations of the $\phi_a$ which correspond to operators with definite conformal dimensions\footnote{We note that, as mentioned below eq. (\ref{unit_norm}), $\delta\t \phi$ defined in eq. (\ref{phiphi_i}) does not appear in the expectation value up to second order in perturbation theory. $\delta\t\phi$ corresponds to the terms with the $C^{(i)}$ and $D^{(i)}$ ($i\in\mathbb Z_{>0}$) coefficients in eq. (\ref{phiphi}) and do not appear at the first and second order amplitudes in eq.s (\ref{bone_i}) and (\ref{bone}), respectively.\label{fn_bone}}.

\b

(e) In our system, we  have states $|\Phi^{(m)}\rangle $  labelled by a parameter $m\in\mathbb Z_{\ge0}$, see eq. (\ref{threeq}).  As we go to higher $m$, the number of states with the same conformal dimensions as $|\Phi^{(m)}\rangle $ increases; in fact even for the lowest interesting value, $m=2$, the number of degenerate states is large enough to make the computation of the matrix $A^{(2)}_{ab}$ difficult. We will be interested in computing something a bit different. The state $|\Phi^{(m)}\rangle $ of interest to us is one of the states $|\phi_a\rangle$; let us call it $|\phi_1\rangle$. Then we compute the quantity $A^{(2)}_{11}$ in eq. (\ref{A2abT}). From (\ref{bone}) we see that the coefficient of $-Te^{-ET}$ in $\hat A^{(2)}$ is
\be
\sum_{a'} |C_{1a'}|^2 \, E^{(2)}_{a'}= \sum_{a'} |\langle \t\phi_{a'}|\phi_1\rangle|^2 E^{(2)}_{a'}\ .
\label{nfour}
\ee
Thus we get the {\it expectation value} of the increase in energy for the  state $|\phi_1\rangle = |\Phi^{(m)}\rangle $.  Computing this quantity will allow us to make our arguments about the nature of lifting of string states.

\subsection{The general expression for lifting at second order}\label{subsec_Aab}
In the above discussion we have expressed the amplitude $A_{ab}$ in eq. (\ref{Aab_define}) in terms of Hamiltonian evolution. But  we will actually compute $A_{ab}$ using path integrals, since the perturbation is known as a change to the Lagrangian rather than a change to the Hamiltonian:
\be
S_0\r S_{\mathrm{pert}}=S_0+\lambda \int d^2 w D(w, \bar w)\ ,
\label{assevent}
\ee
where $D(w,\bar w)$ is an exactly marginal operator deforming the CFT. As mentioned before, $A^{(1)}_{ab}=0$, see the discussion below eq. (\ref{bone_i}). We will work with the next order, where we have
\be
A^{(2)}_{ab}(T)=\h \bigg\langle \phi_b({\tfrac T2})\bigg|\bigg(\int d^2w_2  D(w_2, \bar w_2)\bigg)\bigg(\int d^2w_1  D(w_1, \bar w_1)\bigg)\bigg|\phi_a(-{\tfrac T2})\bigg\rangle\ ,
\label{wnasel}
\ee
where the range of the $w_i$ integrals are
\be
0\le \sigma_i<2\pi\ , ~~~-{\tfrac T 2} <\tau_i<{\tfrac T 2}\ .
\label{nrange}
\ee

Before proceeding, we write the initial and final states in a convenient form. The local operators like $J^a(w)$ in the CFT depend on time $\tau$. Thus if we create the state $|\psi_a\rangle$ by the application of such an operator, then the value of $\tau$ at the point of application is relevant. But if we expand in modes $J^a(w)=\sum_n J^a_n e^{nw}$, then the operators $J^a_n$ do not have the information about the point of application. It is convenient to write the state in terms of mode operators like $J^a_n$, and so we need to factor out the $\tau$-dependence explicitly. 

For $\tau<-{T\over 2}$, the state is the NS vacuum $|0\rangle$. Suppose that the state created at $\tau=-{T\over 2}$ has energy $E$.  Then we have
\be
\big|\phi(-{\tfrac T2})\big\rangle=e^{-{ET\over 2}}|\Phi\rangle\ ,
\ee
where the state $ |\Phi\rangle$ is written with upper case letters: this will denote the fact that this state is made from modes like $J^a_n$ which have no $\tau$-dependence. 
Similarly, the final state  is
\be
\big\langle \phi({\tfrac T2})\big|=e^{-{ET\over 2}} \langle \Phi|\ .
\ee
Eq. (\ref{wnasel}) then reads:
\be
A^{(2)}_{ab}(T)=\h e^{-ET}\bigg\langle \Phi_b({\tfrac T2})\bigg|\bigg(\int d^2w_2  D(w_2, \bar w_2)\bigg)\bigg(\int d^2w_1  D(w_1, \bar w_1)\bigg)\bigg|\Phi_a({-\tfrac T2})\bigg\rangle\ .
\label{wnasel_ii}
\ee

To compute  $A^{(2)}_{ab}$,  we proceed as follows:

\b

(a) Since $\Phi_a$ has the same energy as $\Phi_b$, the integrand depends only on 
\be
\Delta w = w_2-w_1\ .
\ee
It would be convenient if we could write the integrals over $w_1, w_2$ as an integral over $\Delta w$, and factor out the integral over 
\be
s=\h (w_1+w_2)\ .
\ee
We cannot immediately do this, however, as the ranges of the $\tau_i$ integrals given in  (\ref{nrange}) do not factor into a range for $\Delta w$ and a range for $s$. But for our case, we will see that we {\it can} obtain the needed factorization by taking the limit $T\r \infty$. 

Suppose that $w_1<w_2$. In the region $-{T\over 2}<\tau<\tau_1$, we have the state $\Phi_a$, and Hamiltonian evolution gives the factor $\sim e^{-E\tau}$. Similarly, in the region $\tau_2<\tau<{T\over 2}$, we have the state $\Phi_b$, and Hamiltonian evolution  gives $\sim e^{-E\tau}$. In the region $\tau_1<\tau<\tau_2$, we have a state $\Phi_k$ with energy $E_k$, giving a factor $\sim e^{-E_k\tau}$. As we will show below, we have
\be
E_k\ge E+2\ ,
\label{ntwo}
\ee
so that the integrand in $A^{(2)}_{ab}$ in eq. (\ref{wnasel_ii}) is exponentially suppressed as we increase $\Delta w$. Thus  we can fix $w_1=0$, and integrate over $w_2\equiv w$ to compute
\be
\bigg\langle \phi_b({\tfrac T2})\bigg|\bigg(\int d^2w  D(w, \bar w)\bigg) \,  D(0)\bigg|\phi_a(-{\tfrac T2})\bigg\rangle\ .
\label{wnaselqqpre}
\ee
Here the $w$ integral ranges over $0\le \sigma<2\pi, -{T\over 2}<\tau<{T\over 2}$. The $\tau$  range is large  in the limit $T\r\infty$. But the contributions to the integral die off quickly for $|w|\gg 2\pi$. Integration over $w_1$ then just gives a factor
\be
\int d^2 w_1 \r 2\pi T\ .
\ee
Thus in the limit $T\r\infty$, eq. (\ref{wnasel_ii}) reads
\be
 A^{(2)}_{ab}(T)=(2\pi T)\,\h\,e^{-ET}\bigg\langle\Phi_b({\tfrac T2})\bigg|\bigg(\int d^2w  D(w, \bar w)\bigg)\,D(0)\bigg|\Phi_a(-{\tfrac T2})\bigg\rangle\ .
\label{wnaselqq}
\ee
To prove eq. (\ref{ntwo}), we note that $E_k$ must lie in the conformal block of some primary operator $\chi$ with dimensions $(h_\chi, \bar h_\chi)$. Thus, we need a non-vanishing 3-point function
\be
f=\langle \phi_a (z_1) \, D(z_2, \bar z_2) \, \chi(z_3, \bar z_3)\rangle\ .
\ee
Since $\phi_a$ has dimensions $(h,0)$, there is no power of $\bar z_2-\bar z_1$ in the correlator. Since $D$ has dimensions $(h_D,\bar h_D)=(1,1)$, this implies that $\bar h_\chi=1$. Further, since the $w_i$ are integrated over the spatial coordinates $\sigma_i$ with no phase, the state $\phi_k$ must have the same spin as $\phi_a$; i.e. $h_\chi-\bar h_\chi=h$. Thus we have
\be
(h_\chi, \bar h_\chi)=(h+1,1)\ .
\ee
and the lowest state $E_k$ corresponding to such a primary has $E_k=E+2$. If we have a descendent of this lowest state, then we have $E_k>E+2$. Thus we obtain (\ref{ntwo}).

\b

(b) Now consider the integrand of $A^{(2)}_{ab}$ in eq. (\ref{wnaselqq}). We have the correlation function
\be
\langle \Phi_a | \, D(w, \bar w) \, D(0)\, | \Phi_b\rangle\ .
\ee
The right-moving dimensions of $\Phi_a$ and $\Phi_b$ are zero, so the antiholomorphic part of this correlator is $\langle 0 | \, D( \bar w) \,  D(0)\, | 0\rangle$. Since $\bar h_D=1$, we find
\be\label{rightsigmasigma}
\langle 0 | \, D( \bar w) \,  D(0)\, | 0\rangle= {C_1\over  \sinh^2 ({\bar w\over 2})}
\ee
for some constant $C_1$. The left moving part is more complicated and we will calculate it in later sections. But this part also has the same singularity as the right movers when the two $D$ operators approach. So the full correlator will have the form 
\be
\langle \Phi_a | \, D(w, \bar w) \, D(0)\, | \Phi_b\rangle={Q_{ab}(w)\over  \sinh^2({w\over 2})}\,{1\over \sinh^2({\bar w\over 2})}\ .
\label{bbtwo}
\ee
We define
\be
X_{ab}(T)\equiv\int d^2w  {Q_{ab}(w)\over   \sinh^2  ({w\over 2})}\,{1\over  \sinh^2 ({\bar w\over 2})}\ .
\label{wnaselqa}
\ee
The amplitude (\ref{wnaselqq}) then reads
\be
 A^{(2)}_{ab}(T)=\h\,(2\pi T)\,e^{-ET}\,X_{ab}(T)\ .
\label{nthree}
\ee

\begin{figure}
\begin{center}
\includegraphics[width=40mm]{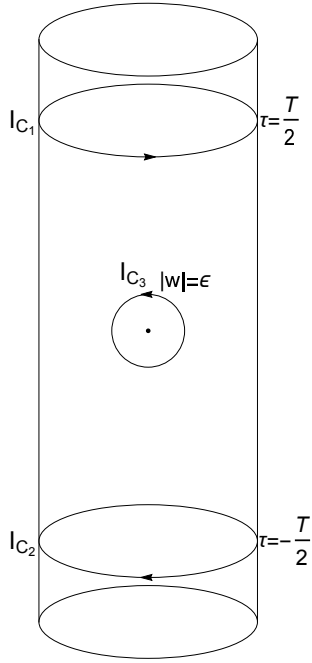}
\end{center}
\caption{Here we show the locations of the three boundary contour integrals, $I_{C_1},I_{C_2}$, and $I_{C_3}$ given in equations (\ref{IC1}), (\ref{IC2}), and (\ref{IC3}) respectively.}
\label{contours}
\end{figure}

\b

(c) To evaluate eq. (\ref{wnaselqa}) we write
\bea
 X_{ab}(T)&=& \int d^2w  {Q_{ab}(w)\over   \sinh^2  ({w\over 2})}\,{1\over  \sinh^2 ({\bar w\over 2})}\nn
&=& \int d^2w  {Q_{ab}(w)\over   \sinh^2  ({w\over 2})}\left ( \p_{\bar w}\coth(\tfrac{\bar w}2)\right ) \nn
&=& -2\int d^2w \, \p_{\bar w} \left (  {Q_{ab}(w)\over   \sinh^2  ({w\over 2})}\,  \coth(\tfrac{\bar w}2)\right ) \nn
&=&  i \int_C dw \,  \left (  {Q_{ab}(w)\over   \sinh^2  ({w\over 2})}\,  \coth({\tfrac{\bar w} 2})\right )\ ,
\label{wnaselqb}
\eea
where in the last line we have used the divergence theorem in complex coordinates. The boundary integral is defined over a contour $C$ consisting of three parts:

\b

(i) $C_1$: the upper boundary of the integration range at $\tau={T\over 2}$. This integral, which we call $I_{C_1}$, runs in the direction of positive $\sigma$. Note that $i(dw)=i(id\sigma)=-d\sigma$ and we have
\be
I_{C_1}=-\int_{0}^{2\pi}  d\sigma \left (  {Q_{ab}(w)\over   \sinh^2  ({w\over 2})}\,  \coth({\tfrac{\bar w} 2})\right )\ .
\label{IC1}
\ee

\b

(ii) $C_2$: the lower boundary of the integration range at $\tau=-{T\over 2}$. The contour runs in the direction of negative $\sigma$. The integral is called $I_{C_2}$ and has the form 
\be
I_{C_2}=\int_{0}^{2\pi}  d\sigma \left (  {Q_{ab}(w)\over   \sinh^2  ({w\over 2})}\,  \coth({\tfrac{\bar w} 2})\right)\ .
\label{IC2}
\ee

\b

(iii) $C_3$: An integral over a small circle of radius $\epsilon$ around the origin where we have the operator $D(0)$. The contour here runs clockwise as this is an inner boundary of the integration domain. We therefore write it as
\be
I_{C_3}=-i\int_{|w|=\epsilon} dw \left (  {Q_{ab}(w)\over   \sinh^2  ({w\over 2})}\,  \coth({\tfrac{\bar w} 2})\right )
\label{IC3}
\ee
where now the integral runs in the usual anticlockwise direction. 
The integral over $C_3$ contains divergent contributions from the appearance of operators with dimension $h+\bar h \le 2$ in the OPE $D(w, \bar w) D(0,0)$. These divergences have to be removed by adding counterterms terms to the action. Thus we get
\be
I_{C_{3}}+I_{C_3,\,\mathrm{counterterm}}= I_{C_3,\,\mathrm{renormalized}}\ .
\ee
Then eq. (\ref{wnaselqb}) reads
\be
X_{ab}(T)=I_{C_1}+I_{C_2}+I_{C_3,\,\mathrm{renormalized}}\ .
\label{bsix}
\ee
Fig. \ref{contours} shows the locations of the three contours.
\b

(iv) Let us now summarize the above discussion. As mentioned in section \ref{outline}(d), we compute $A^{(2)}_{11}\equiv A^{(2)}$ for just one state $|\Phi_1\rangle$, see eq. (\ref{nfour}). This will give the expectation value of the increase in energy of $|\Phi_1\rangle$. From eq.s (\ref{nthree}) and (\ref{bsix}) we obtain
\be\label{A2T_i}
A^{(2)}(T)=\pi T e^{-ET} \left ( I_{C_1}+I_{C_2}+I_{C_3,\,\mathrm{renormalized}}\right )\ .
\ee
Finally, for our state $|\Phi_1\rangle$, the lift in the expectation value of the energy is given by the coefficient of $-Te^{-ET}$ in the limit $T\to\infty$, see eq.s (\ref{bone}) and (\ref{E2T}). Thus, we find
\be
\langle E^{(2)}\rangle=-\pi \lim_{T\r\infty} X_{ab}(T)=-\pi \lim_{T\r\infty}\left(I_{C_1}+I_{C_2}+I_{C_3,\,\mathrm{renormalized}}\right)\ .
\label{deltaE}
\ee

\section{Setting up the computation}\label{section3}
\subsection{The deformation operator}\label{sebsec_deformation}
The orbifold CFT describes the system at its `free' point in moduli space. To move towards the supergravity description, we deform the orbifold CFT by adding a deformation operator $D$, as noted in (\ref{assevent}).

To understand the structure of $D$ we recall that the orbifold CFT contains `twist' operators. Twist operators can link any number $k$ out of the $N$ copies of the CFT together to give a $c=6$ CFT living on a circle of length $2\pi k$ rather than $2\pi$.    We will call such a set of linked copies a `component string' with winding number $k$.

The deformation operator contains a twist of order $2$. The twist itself carries left and right charges $j=\pm \h, \bar j=\pm \h$ \cite{lm2}. Suppose we start with both these charges positive; this gives the twist $\sigma_2^{++}$. Then the deformation operators in this twist sector have the form
\be\label{exactlymarginal}
D=P^{\dot A\dot B}\hat O_{\dot A\dot B}= P^{\dot A \dot B}G^-_{\dot A, -\h}\bar G^-_{\dot B, -\h} \sigma^{++}_2\ .
\ee
Here $P^{\dot A \dot B}$ is a polarization. We will later choose
\be
P^{\dot A \dot B}=\epsilon^{\dot A \dot B}
\ee
where $\epsilon^{+-}=-1$. This choice gives a deformation carrying no charges.

We will omit the subscript $2$ on the twist operator from now on, and will also consider its holomorphic and antiholomorphic parts separately. 
We normalize the twist operator as
\be
\sigma^{-}(z )\sigma^{+}(z')\sim {1\over (z-z')}\ .
\label{asone}
\ee
We note that \cite{Avery:2010er,Avery:2010qw}
\be
G^-_{\dot A, -\h}\sigma^+=-G^+_{\dot A, -\h} \sigma^-\ .
\label{eeone}
\ee
It will be convenient to write one of the two deformation operators as $G^-_{\dot A, -\h}\sigma^+$ and the other as $-G^+_{\dot C, -\h} \sigma^-$. We will make this choice for both the left and right movers, so  the negative sign in (\ref{eeone}) cancels out. Thus on each of the left and right sides we  write one deformation operator in the form $G^-_{\dot A, -\h}\sigma^+$ and the other in the form  $G^+_{\dot A, -\h} \sigma^-$.

From (\ref{asone}) we find that on the cylinder
\be
\langle 0 | \sigma^-(w_2) \sigma^+(w_1)|0\rangle = {1\over 2 \sinh ({\Delta w\over 2})}
\label{axthree}
\ee
where
\be
\Delta w = w_2-w_1\ .
\ee

\subsection{The states}

We start by looking at a CFT with $N=2$; i.e., we have two copies of the $c=6$ CFT (we will consider general values of $N$ in section \ref{section7}). The vacuum $|0\rangle$ with $h=j=0$ is given by two singly-wound copies of the CFT, i.e. there is no twist linking the copies, and the fermions on each of the copies are in the NS sector. Thus we can write
\be
|0\rangle=|0\rangle^{(1)}\,|0\rangle^{(2)}\ ,
\ee
where the superscripts indicate the copy number.

We consider one of the copies to be excited by the application of $m$ R-current operators. The orbifold symmetry requires that the state be symmetric between the two copies, so the state we take is
\bea
|\Phi^{(m)}\rangle&=&{1\over \sqrt{2}}\,
\Big(J^{+(1)}_{-(2m-1)}\dots J^{+(1)}_{-3}J^{+(1)}_{-1} ~+~  J^{+(2)}_{-(2m-1)}\dots J^{+(2)}_{-3}J^{+(2)}_{-1}\Big)|0\rangle\nn
&\equiv& |\Phi^{(m)}\rangle^{(1)}~+~|\Phi^{(m)}\rangle^{(2)}\ ,
\label{threex}
 \eea
where in $|\Phi^{(m)}\rangle^{(i)}$ the excitations act on copy $i$. This state has
\be
(h,\bar h)=(m^2,0),\qquad(j,\bar j)=(m,0)\ .
\ee
The energy of the state is 
\be
E\equiv h+\bar h  = m^2
\label{assix}
\ee
and its momentum is 
\be
P\equiv h-\bar h =m^2\ .
\label{asfive}
\ee
The final state is the conjugate of the initial state
\be
\langle \Psi^{(m)}|={1\over \sqrt{2}}\;\langle 0|\Big( J^{-(1)}_{1}J^{-(1)}_{3}\dots J^{-(1)}_{(2m-1)}+
J^{-(2)}_{1}J^{-(2)}_{3}\dots J^{-(2)}_{(2m-1)}\Big)\ . 
\label{threexq}
\ee

\section{The vacuum correlator}\label{section4}

\begin{figure}
\begin{center}
\includegraphics[width=41mm]{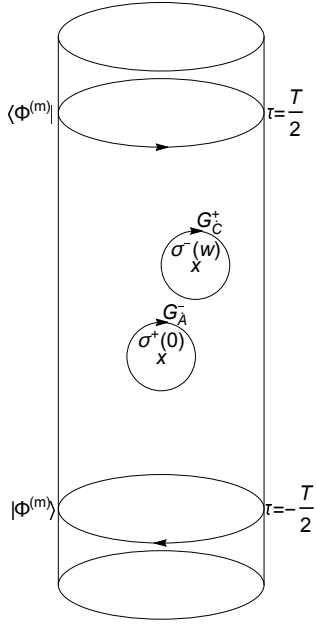}
\end{center}
\caption{The cylinder with the locations of the initial state $|\Phi^{(m)}\rangle$ at $\tau = -{T\over2}$, the final state $\langle\Phi^{(m)}|$ at $\tau={T\over2}$, and the two deformation operators  $G^-_{\dot{A}}\sigma^+$ and $G^+_{\dot{C}}\sigma^-$ at $w_1=0$ and $w_2=w$, respectively.}
\label{cylinderstate}
\end{figure}

As a first step, we compute the vacuum to vacuum correlator
\be
T_{\dot C\dot A}(w_2,w_1)=\langle0|\big(G^+_{\dot C, -\h}\sigma^-(w_2)\big)\,\big(G^-_{\dot A, -\h}\sigma^+(w_1)\big)\big|0\rangle\ .
\label{adoneq}
\ee
The complex conjugate of this correlator will give the right moving part of the correlator of $A^{(2)}(T)$, see eq. (\ref{wnaselqq}). Fig.\ref{cylinderstate} represents the full state on the cylinder.

\subsection{The map to the covering space}\label{map}

To compute the vacuum amplitude $T_{\dot C\dot A}(w_2,w_1)$ in eq. (\ref{adoneq}) we first map the cylinder labeled by $w$ to the complex plane labeled by $z$:
\be
z=e^w\ .
\ee
We then map this plane to its covering space where the twist operators are resolved, see \cite{lm1} for the details of the covering space analyses. We consider the map
 \be
z={(t+a)(t+b)\over t}\ .
\label{amap}
\ee
We have
\be
{dz\over dt} = 1-{ab\over t^2}\ .
\ee
The twist operators correspond to the locations given by ${dz\over dt}=0$; i.e. the points
\bea
&&t_1=-\sqrt{ab}\ ,\qquad z_1=e^{w_1}=(\sqrt{a}-\sqrt{b})^2\ ,\\
&&t_2=\sqrt{ab}\ ,\qquad\;\;\, z_2=e^{w_2}=(\sqrt{a}+\sqrt{b})^2\ .
\eea
Note that
\be
{dz\over dt} = {(t-t_1)(t-t_2)\over t^2}\ .
\ee
We define
\be
\Delta w = w_2-w_1\ ,
\ee
\be
s={\h}(w_1+w_2)\ ,
\ee
Then we find
\be
a=  e^{s} \cosh^2 (\tfrac{\Delta w }{4}), ~~~b=  e^{s} \sinh^2 (\tfrac{\Delta w }{4})\ .
\ee
It will be useful to note the relations
\be
a-b=e^{s}, ~~~z_1z_2=e^{2s}, ~~~z_1-z_2= -2 e^{s} \sinh (\tfrac{\Delta w }{2})\ .
\ee

\subsection{The `base' amplitude}\label{base}

To compute the vacuum correlator $T_{\dot C\dot A}(w_2,w_1)$ in eq. (\ref{adoneq}), we start by computing
\be
U(w_2,w_1)\equiv\langle 0|  \sigma^-(w_2)\,  \sigma^+(w_1) |0\rangle\ .
\label{axone}
\ee
We call this the `base' amplitude since each correlator we compute will have this structure of twist operators, and the only extra elements will be local operators with no twist. 

The computation of correlators like (\ref{axone}) is discussed in \cite{lm1,lm2}. We briefly summarise the computation by proceeding in the following steps:

\b

(i) We already mapped the cylinder labelled by the coordinate $w$ to the complex plane labelled by the coordinate $z$ in subsection \ref{map}, through the map $z=e^w$. We then mapped the plane to the covering space, labelled by the coordinate $t$, though the map (\ref{amap}). These maps generate a Liouville factor since the curvature of the covering space is different from the curvature on the cylinder, and the change of curvature changes the partition function due to the conformal anomaly of the CFT. Let this Liouville factor be
\be
L\,[z_1,z_2]\ .
\label{axtwo}
\ee

\b

(ii) The twist operator $\sigma^+(w_1)$ has left-moving dimension $h=\h$ and transforms to the plane as
\be\label{sigmaplus}
\left ( {dz\over dw}(z_1)\right ) ^\h  \sigma^+(z_1)=z_1^\h \sigma^+(z_1)\ .
\ee
Now consider the map to the cover. A twist $\sigma^+(z=0)$ on the plane $z$ transforms to a spin field ${S}^+(t=0)$ on the covering space\footnote{This is because fermionic fields have different boundary conditions in the odd versus even twisted sectors of the symmetric orbifold. The NS sector fermions have the usual NS-type half-integer modes in the odd twisted sector, whereas in the even twisted sector they have Ramond-type integer modes. The spin fields account for the ground state energy in the Ramond (R) sector, see \cite[section 2.2 ]{lm2} for details.\label{fn_RNS}} under the map $z=t^2$. The spin field has left-moving dimension $h={1\over 4}$. In the map (\ref{amap}) we have
\be
z-z_1\approx C(t-t_1)^2\ ,\qquad C=-{1\over \sqrt{ab}}\ ,
\ee 
so we have an extra scaling factor $\sqrt{C}$ furnishing $t$ compared to the standard map $z-z_1\approx (t-t_1)^2$, see \cite{lm2}. Combining with eq. (\ref{sigmaplus}), we find that the twist 2 operator $\sigma^+(w_1)$ transforms to the spin field ${S}^+(t_1)$ on the covering surface with an overall factor
\be
z_1^\h\big(\sqrt{C}\big)^{1\over 4}=z_1^\h\left ( -{1\over \sqrt{ab}}\right )^{1\over 8}\ .
\ee

\b

(iii) Similarly, the operator $\sigma^-(w_2)$ transforms to the spin field ${S}^-(t_2)$ on the cover acquiring an overall factor
\be
z_2^\h \left ( {1\over \sqrt{ab}}\right )^{1\over 8}\ .
\ee

\b

(iv) At this stage we have on the $t$ plane the amplitude
\be
\langle0|{S}^-(t_2) \, {S}^+(t_1)|0\rangle\ .
\ee
As discussed in footnote \ref{fn_RNS}, the spin field ${S}^+(t_1)$ creates an R vacuum at $t_1$. We can make a spectral flow transformation around the point $t=t_1$ to map this R vacuum to the NS vacuum $|0\rangle$, see appendix \ref{app_sf} for a brief review.  The NS vacuum is equivalent to no insertion on the covering space at all, so we would have taken all the effects of the twist into account. The spectral flow parameter needed is $\alpha=-1$, and we obtain
\be
{S}^+(t_1)|0\rangle \longmapsto |0\rangle\ .
\ee
Under such a spectral flow transformation, other fields in the $t$ plane pick up a factor as given in eq. (\ref{sf_O}). Thus, the field ${S}^-(t_2)$ (which has R-charge $j=-\h$) acquires a factor $(t_2-t_1)^{-{1\over 2}}$.

\b

(v) Now consider the spin field ${S}^-(t_2)$. We perform a similar spectral flow around the point $t=t_2$ with $\alpha=1$. This gives
\be
{S}^-(t_2)|0\rangle \longmapsto |0\rangle\ .
\ee
There are no other fields in the $t$ plane, so this time we get no additional factors from the spectral flow. 

\b

(vi) We now just have the $t$ plane with no insertions. The amplitude for this vacuum state is unity: it has been set to this value when defining the Liouville factor (\ref{axtwo}). Collecting all the factors (i)-(v) above, we obtain the amplitude $U(w_2,w_1)$ in eq. (\ref{axone}). 

\b

While we can compute $U(w_2,w_1)$ as outlined above, it turns out that we do not need to carry out these steps in this specific example: we already know the result from eq. (\ref{axthree})
\be
U(w_2,w_1)={1\over 2 \sinh({\Delta w\over 2})}\ .
\label{base_amplitude}
\ee
The reason we do not have to carry out the steps (i)-(v) explicitly here is that we have only two twist operators in our correlator; in this situation the factors from steps (i)-(v) can be absorbed in the normalization of the twists. However, if we have more than two twists then we do need to compute all factors explicitly.

Even though we can compute $U(w_2,w_1)$ without carrying out these steps, it is important to list the steps since when we have other excitations in the correlator then we will get additional factors from each of these steps. For later use, it will be helpful to also write the base amplitude (\ref{base_amplitude}) in alternative ways using the relations in section \ref{map}:
\be
U={z_1^\h z_2^\h\over (z_2-z_1)}={(a-b)\over 4 \sqrt{ab}}\ .
\label{bbone}
\ee

\subsection{The complete vacuum amplitude}\label{sect}

We now return to the computation of the vacuum amplitude $T_{\dot C\dot A}(w_2,w_1)$ defined in (\ref{adoneq}). Consider the operator
\be
 G^-_{\dot A, -\h} ={1\over 2\pi i} \int_{w_1} dw'_1 G^-_{\dot A}(w'_1)\ .
 \ee
 We proceed in the following steps:
 
 \b
 
 (i) We have
\bea
 {1\over 2\pi i} \int_{w_1} dw'_1 G^-_{\dot A}(w'_1)&=&{1\over 2\pi i} \int_{z_1} dz'_1 \Big({dz'_1\over dw'_1}\Big)^\h G^-_{\dot A}(z'_1)\\
 &=&{1\over 2\pi i} \int_{t_1} dt'_1\Big({dt'_1\over dz'_1}\Big)^\h\Big({dz'_1\over dw'_1}\Big)^\h G^-_{\dot A}(t'_1)\nn  
 &=&{1\over 2\pi i} \int_{t_1} dt'_1 (t'_1-t_1)^{-\h} (t'_1-t_2)^{-\h} {t'_1}^\h (t'_1+a)^\h (t'_1+b)^\h G^-_{\dot A}(t'_1)\ .\nonumber
\eea

\b

(ii) In section \ref{base} above we have seen that we perform a spectral flow around $t=t_1$ by $\alpha=-1$ and another one around $t=t_2$ by  $\alpha=1$. These flows give the factors
\be
G^-_{\dot A}(t'_1)\longmapsto(t'_1-t_1)^{-\h}(t'_1-t_2)^{\h} G^-_{\dot A}(t'_1)\ ,
\ee
see appendix \ref{app_sf}. Thus, we obtain
\bea
 {1\over 2\pi i} \int_{w_1} dw'_1 G^-_{\dot A}(w'_1) &\longmapsto& {1\over 2\pi i} \int_{t_1} dt'_1 (t'_1-t_1)^{-1} {t'_1}^\h (t'_1+a)^\h (t'_1+b)^\h G^-_{\dot A}(t'_1)\nn
 &=&\!\!\!{t_1}^\h (t_1+a)^\h (t_1+b)^\h G^-_{\dot A}(t_1)\ .
 \eea
 
\b

(iii) Similarly we have
\be
 {1\over 2\pi i} \int_{w_2} dw'_2 G^+_{\dot C}(w'_2)\longmapsto{t_2}^\h (t_2+a)^\h (t_2+b)^\h G^+_{\dot C}(t_2)\ .
 \ee
 
 \b
 
 (iv) Apart from the c-number factors in steps (i)-(iii), we have the $t$ plane correlator
 \be
\langle0|G^+_{\dot C}(t_2)\, G^-_{\dot A}(t_1) |0\rangle~=~\epsilon_{\dot C\dot A}\,{(-2)\over (t_2-t_1)^3}\ .
 \label{axtw}
 \ee
 
\b

(v) Collecting all the factors and noting the base amplitude (\ref{base_amplitude}), we find\footnote{Since there are fractional powers in the expressions here, the overall phase involves a choice of branch. But similar fractional powers appear in the right moving sector, and we can choose the signs as taken here with the understanding that we choose similar signs for the right movers.}
\bea
&&T_{\dot C\dot A}(w_2,w_1)=\left ( {t_1}^\h (t_1+a)^\h (t_1+b)^\h \right )\left ( {t_2}^\h (t_2+a)^\h (t_2+b)^\h \right )\times\nonumber\\
&&\qquad\qquad\qquad\qquad\qquad\qquad\qquad\qquad\qquad\times\left ( \epsilon_{\dot C\dot A}\,{(-2)\over (t_2-t_1)^3}U(w_2,w_1) \right ) \nn
&&\qquad\qquad\quad\,\,=\epsilon_{\dot C\dot A}\,{(a-b)^2\over 16\,ab} \, = \,\epsilon_{\dot C\dot A}\,{1\over 4\sinh^2 ({\Delta w \over 2} )}\ .
\label{axeight}
\eea
The right moving part of the correlator in the integrand of $A^{(2)}(T)$ in eq. ({\ref{A2T_i}) is found by taking the complex conjugate of this expression and taking $\e_{\dot{C}\dot{A}}\to\e_{\dot{D}\dot{B}}$
\be
\big\langle0|\big(G^+_{\dot D, -\h}\sigma^-(\bar w_2)\big)~\big( G^-_{\dot B, -\h} \sigma^+(\bar w_1)\big)\big|0\rangle=
\epsilon_{\dot D\dot B}\,{1\over 4\sinh^2 ({\Delta \bar w \over 2} )}\ .
\ee

\section{Lifting of D1-D5-P states}\label{section5}

In this section we evaluate lifting of the D1-D5-P states (\ref{threex}). We compute the left part of the correlator appearing in the amplitude $A^{(2)}(T)$, see eq.s (\ref{wnaselqq}) and (\ref{A2T_i}). Analogous to (\ref{adoneq}), we define
\be
T^{(j)(i)}_{\dot C\dot A,m}(w_2,w_1)={}^{(j)}\big\langle\Phi^{(m)}\big|\big(G^+_{\dot C, -\h} \sigma^-(w_2)\big)\big(G^-_{\dot A, -\h}\sigma^+(w_1)\big)\big|\Phi^{(m)}\big\rangle^{(i)}\ ,
\label{adoneqq}
\ee
where the superscripts $(i),(j)$ indicate which of the two copies carries the current excitations.

\subsection{Computing $T^{(1)(1)}_{\dot C\dot A,m}(w_2,w_1)$}\label{spectralflowfactors}

Let us start by computing $T^{(1)(1)}_{\dot C\dot A,m}(w_2,w_1)$. We will see that this computation will automatically extend to yield all the $T^{(j)(i)}_{\dot C\dot A,m}(w_2,w_1)$ amplitudes.

The operator
\be
{\cal J}^{+,(m)}\equiv J^{+}_{-(2m-1)}\cdots J^{+}_{-3}J^{+}_{-1} 
\label{calj}
\ee
has quantum numbers $(h,\bar h)=(m^2,0)$ and $(j,\bar j)=(m,0)$. With the commutation relations given in (\ref{commutations_ii}), we find that ${\cal J}^+_m(z=0)$ generates a state with unit norm at $z=0$. The operator conjugate to ${\cal J}^+_m$ is
\be
{\cal J}^{-,(m)}\equiv J^{-}_{1}J^{-}_{3} \cdots J^{-}_{(2m-1)}\ .
\ee

We follow the same process by which we computed the amplitude $T_{\dot C\dot A}(w_2,w_1)$ in subsection \ref{sect}.  The initial and final states have been written in terms of operator modes and can therefore  be assumed to be placed at $\tau\r-\infty$ and $\tau\r \infty$, respectively. We first map the cylinder $w$ to the plane $z$. The currents in the initial state give the operator ${\cal J}^{+,(m)}(z=0)$ on copy $1$. The currents in the final state give the operator ${\cal J}^{-,(m)}(z=\infty)$, again on copy $1$.

Next we map to the $t$ plane via the map (\ref{amap}). The point $z=0$ for copy $1$ maps to $t=-a$. Thus, ${\cal J}^{+,(m)}(z=0)$ maps as
\be
{\cal J}^{+,(m)}(z=0) ~\longmapsto~ \Big({dt\over dz}\Big)^{m^2} {\cal J}^{+,(m)}(t=-a)=\Big({a\over a-b}\Big)^{m^2} {\cal J}^{+,(m)}(t=-a)\ .
\label{axtenq}
\ee
The point $z=\infty$ for copy $1$ maps to $t=\infty$. We have $z\sim t$ at $t=\infty$, so  the operator ${\cal J}^{-,(m)}(z=\infty)$ maps as
\be
{\cal J}^{-,(m)}(z=\infty) ~\longmapsto~ {\cal J}^{-,(m)}(t=\infty)\ .
\label{axelq}
\ee

We note that the state ${\cal J}^{+,(m)}|0\rangle$ is obtained by spectral flow of the vacuum $|0\rangle$ by $\alpha=2m$, see appendix \ref{app_sf}. Thus we can use the spectral flow by $\alpha=-2m$ to map ${\cal J}^{+,(m)}|0\rangle$ to the vacuum $|0\rangle$. We will use this trick to remove the insertion of ${\cal J}^{+,(m)}$ on the $t$ plane: this reduces the amplitude to the one we had for $T_{\dot C\dot A}(w_2,w_1)$ in eq. (\ref{axeight}). (Note that when we remove ${\cal J}^{+,(m)}$ from any point in the $t$ plane, we remove at the same time the operator ${\cal J}^{-,(m)}$ at infinity.)
\begin{figure}
\begin{center}
\includegraphics[width=75mm]{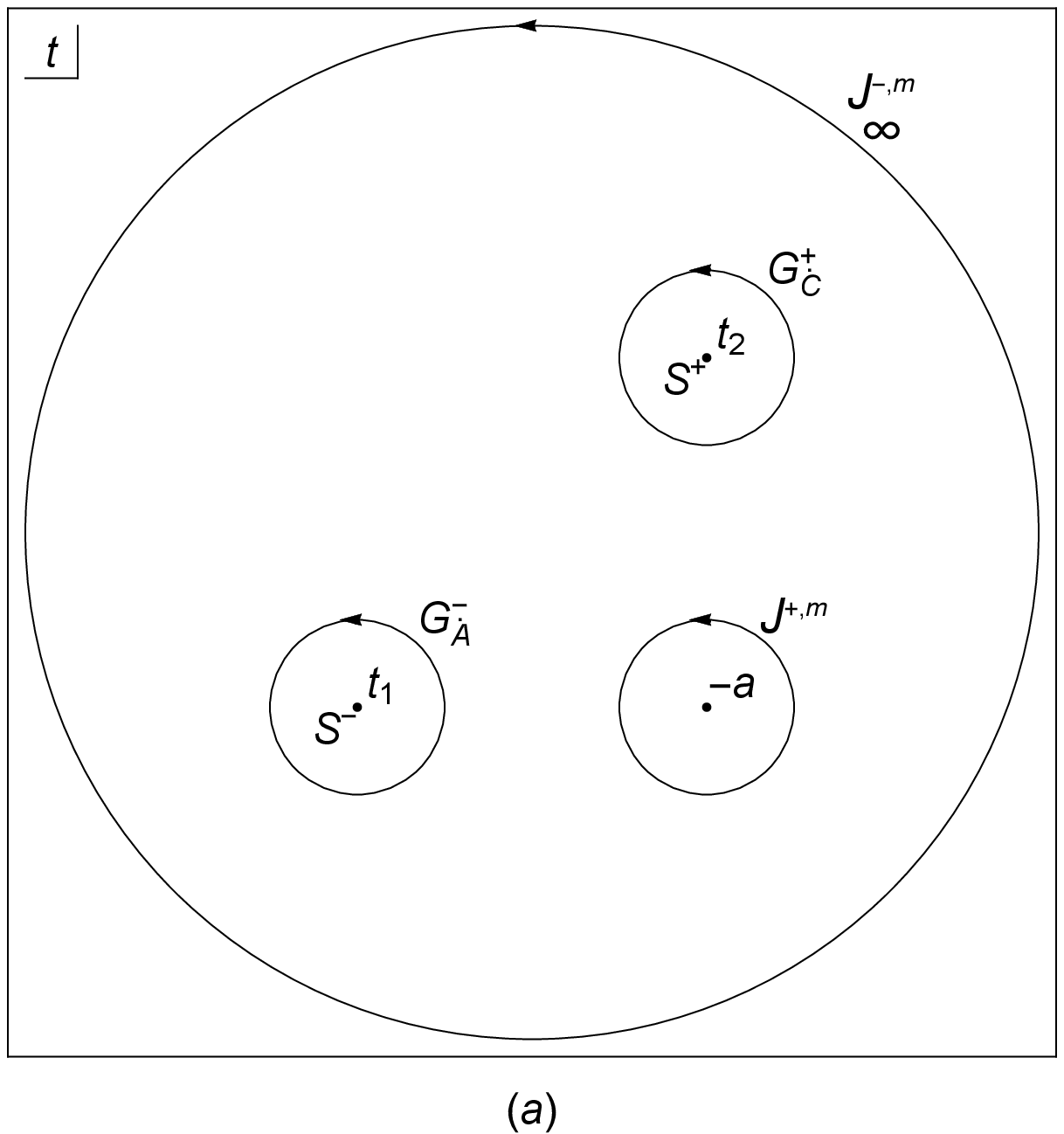}
\vspace{0.3cm}
\hspace{3mm}
\includegraphics[width=75mm]{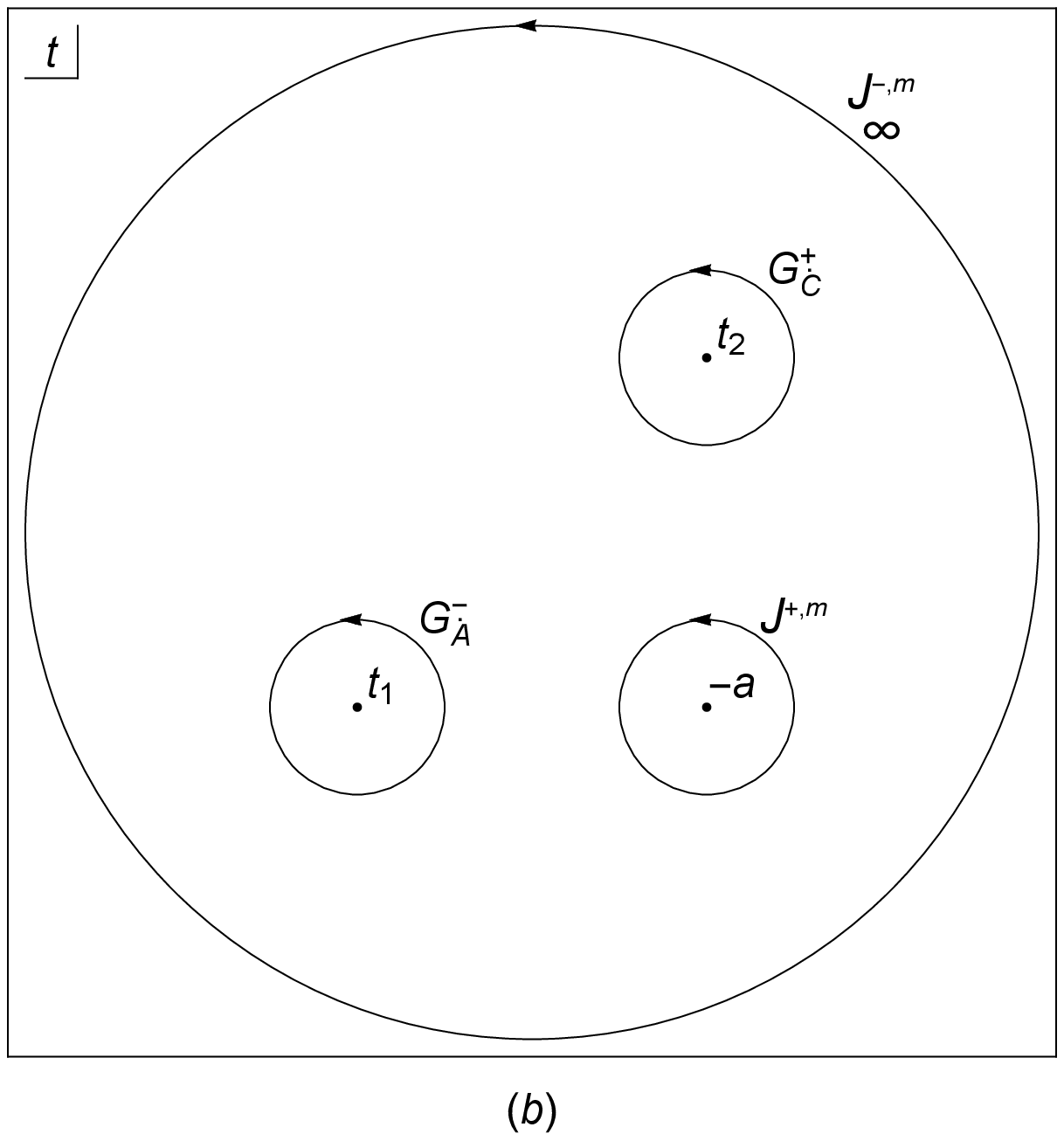}
\includegraphics[width=75mm]{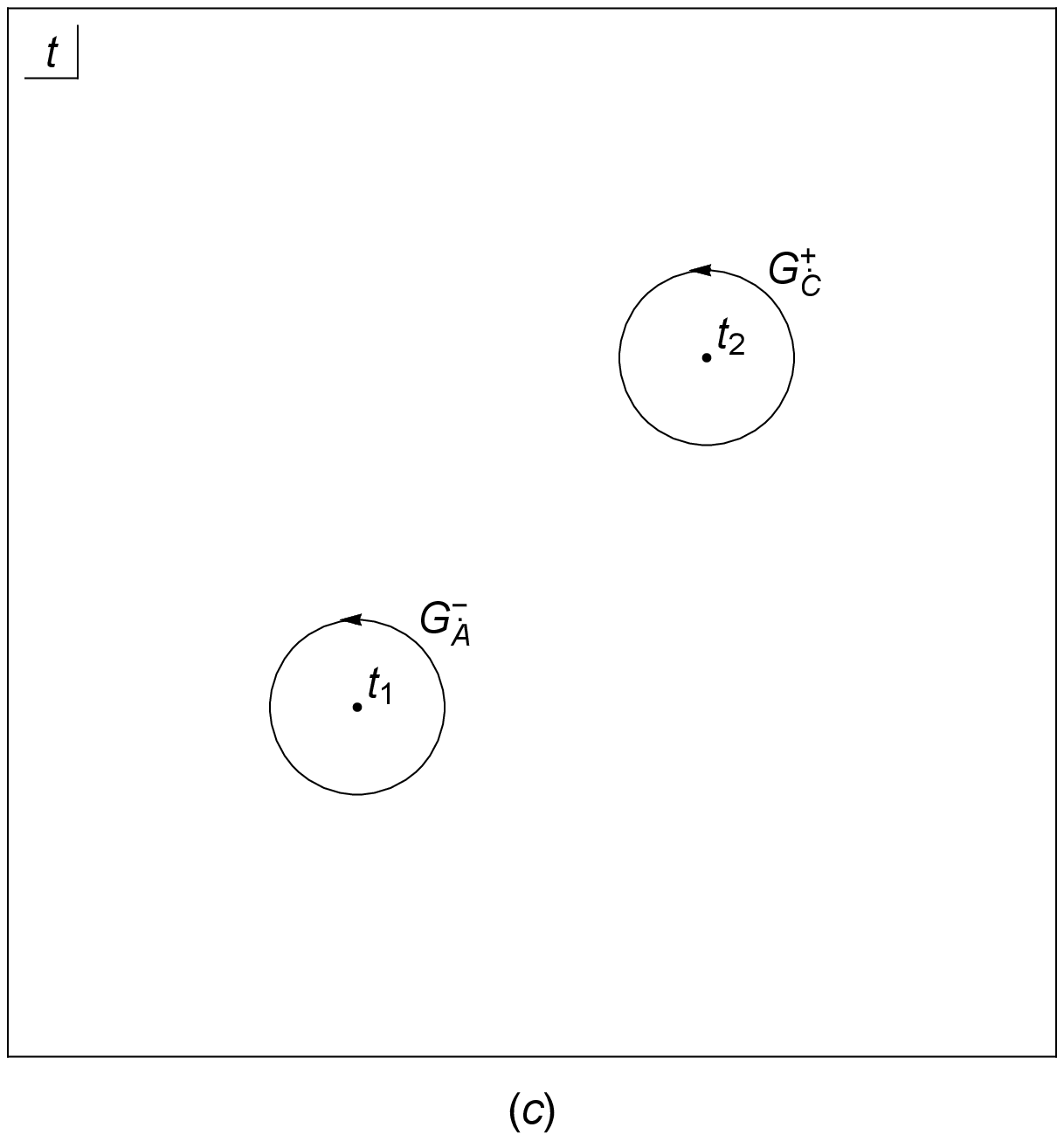}
\end{center}
\caption{(a) The $t$ plane with the spin fields $S^{+}(t_1)$ and $S^{-}(t_2)$, the $G^{-}_{\dot{A}}$ contour circling $t_1$, the $G^{+}_{\dot{C}}$ contour circling $t_2$, the $\mathcal{J}^{+,m}$ contour circling $t=-a$, and the $\mathcal{J}^{-,m}$ contour circling $t=\infty$. (b) The spin fields, $S^{+}(t_1)$ and $S^{-}(t_2)$ spectral flowed away. (c) The spin fields $S^{+}(t_1)$ and $S^{-}(t_2)$ and the currents $\mathcal{J}^{+,m}$ and $\mathcal{J}^{-,m}$ all spectral flowed away. All of the spectral flow factors are given in subsection \ref{spectralflowfactors}.}
\label{fig_spectralflow}
\end{figure}
Fig.\ref{fig_spectralflow} shows the covering space insertions for the unspectral flowed amplitude, the amplitude with only spin fields spectral flowed away, and the amplitude with both spin fields and currents spectral flowed away.

Let us note the extra factors we get in computing $T^{(1)(1)}_{\dot C\dot A,m}(w_2,w_1)$ in eq. (\ref{adoneqq}) as compared to $T_{\dot C\dot A}(w_2,w_1)$ in eq. (\ref{axeight}):

\b

(i) From (\ref{threex}) we see that the initial and final states $|\Psi^{(m)}\rangle^{(1)}$ and  ${}^{(1)}\langle \Psi^{(m)}|$ have a normalization ${1\over \sqrt{2}}$ each; this gives a factor
\be\label{eq_f1}
f_1=\left ( {1\over \sqrt{2}}\right )^2\ .
\ee

(ii) We have the factor  obtained in (\ref{axtenq}) when mapping  ${\cal J}^{+,(m)}$ from the $z$ to the $t$ plane:
\be\label{eq_f2}
f_2=\left({a\over a-b}\right)^{m^2}\ . 
\ee

\b

(iii) We perform a spectral flow by $\alpha=-1$ around the point $t=t_1$. Under this spectral flow, the operator ${\cal J}^{+,(m)}(t=-a)$ picks up the factor
\be\label{eq_f3}
f_3=(-a-t_1)^{m}\ .
\ee

\b

(iv) We perform a spectral flow by $\alpha=1$ around the point $t=t_2$. Under this spectral flow, the operator ${\cal J}^{+,(m)}(t=-a)$ picks up the factor
\be\label{eq_f4}
f_4=(-a-t_2)^{-m}\ .
\ee

\b

(v) We perform a spectral flow around $t=-a$ by $\alpha=-2m$. This gives
\be
{\cal J}^{+,(m)}|0\rangle_{(t=-a)}~\longmapsto~|0\rangle_{(t=-a)}\ .
\ee
We have the operator $G^-_{\dot A}(t=t_1)$; this picks up a factor
\be\label{eq_f5}
f_5=(t_1+a)^{-{m}}\ .
\ee
Likewise, the operator $G^+_{\dot C}(t=t_2)$ picks up a factor
\be\label{eq_f6}
f_6=(t_2+a)^{{m}}\ .
\ee

\b

We are now left with just the amplitude (\ref{axeight}). Combining eq.s (\ref{eq_f1})-(\ref{eq_f6}), we find 
\bea
&&T^{(1)(1)}_{\dot C\dot A,m}(w_2,w_1)=\prod_{i=1}^6f_i\,T_{\dot C\dot A}(w_2,w_1)={\epsilon_{\dot C\dot A}\over 2}\Big({a\over a-b}\Big)^{m^2}  \left [ {(a-b)^2\over 16 ab}\right ] \nn
&&\qquad\qquad\qquad\;=\epsilon_{\dot C\dot A} {\big(\cosh({\Delta w\over 4})\big)^{2m^2}\over 8 \sinh^2 ({\Delta w \over 2})}\ .
\label{axthir}
\eea

\subsection{Computing the remaining $T_{\dot C\dot A}^{(j)(i)}$}

Suppose we fix $\sigma_1$ and consider the shift $\sigma_2\mapsto \sigma_2+2\pi$. This gives $\Delta w \mapsto \Delta w + 2 \pi i$. Under this change $\sinh ({\Delta  w \over 2}) \mapsto -\sinh ({\Delta  w \over 2})$, and so $\sinh^2 ({\Delta  w \over 2})$ is invariant. Similarly, $\sinh^2 ({ \Delta  \bar w \over 2})$ is invariant. But $\cosh({\Delta w\over 4})\mapsto i\sinh({\Delta w\over 4})$. Thus, the integrand in (\ref{axthir}) is not periodic under $\sigma_2\mapsto \sigma_2+2\pi$. The reason is that when we move $\sigma_2$ through $2\pi$, we move from copy $1$ to copy $2$. This implies
\be
T_{\dot C\dot A, m}^{(1)(1)}~\mapsto~ T_{\dot C\dot A,m}^{(2)(1)}\ .
\ee
Under the shift $\sigma_2\mapsto \sigma_2+2\pi$ we find
\be
\big(\cosh(\tfrac{\Delta w}{4})\big)^{2m^2}\mapsto (-1)^m\big(\sinh(\tfrac{\Delta w}{4})\big)^{2m^2}\ .
\ee

Thus, we see that we can take into account all the four terms $T^{(j)(i)}$ by taking $T^{(1)(1)}$ and making the replacement
\be
\big(\cosh(\tfrac{\Delta w}{4})\big)^{2m^2}\mapsto 2\left ( \big(\cosh(\tfrac{\Delta w}{4})\big)^{2m^2}+(-1)^m\big(\sinh(\tfrac{\Delta w}{4})\big)^{2m^2}\right )\ .
\ee
Collecting the left and right parts of the correlator, we find that
\be
\langle \Psi^{(m)}| D(w) D(0)|\Psi^{(m)}\rangle = P^{\dot A\dot B} P^{\dot C \dot D} \epsilon_{\dot C\dot A}\epsilon_{\dot D\dot B} {\left ( \big(\cosh({\Delta w\over 4})\big)^{2m^2}+(-1)^m\big(\sinh({\Delta w\over 4})\big)^{2m^2}\right )\over 16 \sinh^2 ({\Delta w \over 2}) \sinh^2 ({\Delta \bar w \over 2})}\ .
\ee
Comparing with (\ref{bbtwo}), we find that
\be
Q^{(m)}(w)=P^{\dot A\dot B} P^{\dot C \dot D} \epsilon_{\dot C\dot A}\,\epsilon_{\dot D\dot B}\,{\left ( \big(\cosh({\Delta w\over 4})\big)^{2m^2}+(-1)^m
\big(\sinh({\Delta w\over 4})\big)^{2m^2}\right )\over 16 }\ .
\label{bbthree}
\ee

\subsection{Computing $X^{(m)}(T)$ for $m$ even}

Due to the term $(-1)^m$ in (\ref{bbthree}), it is convenient to treat the cases of even and odd $m$ separately. We consider even values of $m$ in this subsection and treat the odd $m$ case in the next subsection.

We first compute the contour integral $I_{C_1}$ in eq. (\ref{IC1}) in the limit $\tau\r\infty$. To do so, we set $w_2\equiv w$, $w_1=0$, and expand the functions in (\ref{bbthree}) in powers of $e^{-w}$. We find:
\bea
\big(\cosh(\tfrac{w}{4})\big)^{2m^2}&=&{1\over 2^{2m^2}} e^{{m^2\over 2} w}\sum_{k=0}^{2m^2}\, {}^{2m^2}C_k e^{-{k\over 2}w}\ ,\nn
\big(\sinh(\tfrac{w}{4})\big)^{2m^2}&=&{1\over 2^{2m^2}} e^{{m^2\over 2} w}\sum_{k=0}^{2m^2}\, {}^{2m^2}C_k (-1)^k e^{-{k\over 2}w}\ ,
\eea
where ${}^mC_n$ are the binomial coefficients. Defining $k=2k'$, $k^\prime\in\mathbb Z$, we find
\be
\big(\cosh(\tfrac{w}{4})\big)^{2m^2}+\big(\sinh(\tfrac{w}{4})\big)^{2m^2}={2\over 2^{2m^2}} e^{{m^2\over 2} w}\sum_{k'=0}^{m^2}\, {}^{2m^2}C_{2k'}  e^{-k'w}\ .
\ee
We also have
\bea
&&{1\over  \sinh^2 ({w\over 2})}=4e^{-w}\sum_{l=0}^\infty (l+1) e^{-l w}\ ,\\
&&\coth(\tfrac{\bar w}{2})=(1+e^{-\bar w})\sum_{n=0}^\infty  e^{-n\bar w}\ .
\eea
We have $I_{C_1}$ in eq. (\ref{IC1}) as an integral over $w$ at $\tau={T\over 2}$:
\bea
I_{C_1}&=&-P^{\dot A \dot B}P^{\dot C\dot D} \epsilon_{\dot C\dot A}\epsilon_{\dot D\dot B}\cr
&&~~~~~~~~\times ~~{1\over 2}\int_{\sigma=0}^{2\pi} d\sigma  \, \left ( {e^{{m^2\over 2} w}\over 2^{2m^2}} \right )
\left ( \sum_{k'=0}^{m^2}\, {}^{2m^2}C_{2k'}  e^{-k'w}\right ) \left (e^{-w}\sum_{l=0}^\infty (l+1) e^{-l w}\right )\times\nn
&&
~~~~~~~~~~~~~~~\times ~~ \left ((1+e^{-\bar w})\sum_{n=0}^\infty  e^{-n\bar w} \right ) \ ,
\label{mone_i}
\eea
where the last bracket contains antiholomorphic factors of the form  $1, e^{-\bar w}, e^{-2\bar w},\cdots$. We will now argue that only the leading term, $1$, survives from this bracket, in the limit $T\r \infty$. To see this, note that the first bracket on the RHS of (\ref{mone_i}) has a power $e^{{m^2\over 2} w}$. From the second and third brackets, we can get a power $e^{-k_1w}$ with $k_1\ge 0$ and from the last bracket we can get a power $e^{-k_2 \bar w}$ with $k_2\ge 0$. These factors give
\be
e^{({m^2\over 2}-k_1-k_2)\tau} e^{i({m^2\over 2}-k_1+k_2)\sigma}\ .
\ee
We have
\be
\int_0^{2\pi} d\sigma e^{i({m^2\over 2}-k_1+k_2)\sigma}= 2\pi \delta_{{m^2\over 2}-k_1+k_2,0}\ ,
\label{mtwo}
\ee
so that we have $k_1={m^2\over 2}+k_2$. The power of $e^{\tau}$ then gives (since $\tau={T\over 2}$)
\be
e^{-k_2{T}}\ .
\ee
Thus in the limit $T\r\infty$, the only surviving term is $k_2=0$, and therefore the last bracket in (\ref{mone_i}) can be replaced by unity. We then get 
\be
k_1=k'+l+1={m^2\over 2}
\ee
which sets $l={m^2\over 2} -k'-1$. The condition $l\ge 0$ then gives
\be
k'\le {m^2\over 2}-1\ .
\ee
Using (\ref{mtwo}), we find
\bea
I_{C_1}&=&-P^{\dot A \dot B}P^{\dot C\dot D} \epsilon_{\dot C\dot A}\epsilon_{\dot D\dot B}\,{\pi\over 2^{2m^2}}\sum_{k'=0}^{{m^2\over 2}-1} \, {}^{2m^2}C_{2k'} \,(\tfrac{m^2}2-k')\cr
&=& -P^{\dot A \dot B}P^{\dot C\dot D} \epsilon_{\dot C\dot A}\epsilon_{\dot D\dot B}\,{\sqrt{\pi} \over 8}\,{\Gamma[m^2-\h]\over \Gamma[m^2-1]}\ .
\eea

We proceed similarly for the contour integral $I_{C_2}$ in eq. (\ref{IC2}) at $\tau=-{T\over 2}$. This time we expand the functions in (\ref{bbthree}) in powers of $e^w$:
\bea
\big(\cosh(\tfrac{w}{4})\big)^{2m^2} &=&{1\over 2^{2m^2}} e^{-{m^2\over 2} w}\sum_{k=0}^{2m^2}\, {}^{2m^2}C_k e^{{k\over 2}w}\ ,\nn
\big(\sinh(\tfrac{w}{4})\big)^{2m^2}&=&{1\over 2^{2m^2}} e^{-{m^2\over 2} w}\sum_{k=0}^{2m^2}\, {}^{2m^2}C_k (-1)^k e^{{k\over 2}w}\ .
\eea
Defining again $k=2k'$, $k'\in\mathbb Z$ in these sums, we find
\be
\big(\cosh(\tfrac{w}{4})\big)^{2m^2}+\big(\sinh(\tfrac{w}{4})\big)^{2m^2}={2\over 2^{2m^2}} e^{-{m^2\over 2} w}\sum_{k'=0}^{m^2}\, {}^{2m^2}C_{2k'}  e^{k'w}\ .
\ee
We further have
\bea
{1\over  \sinh^2 (\tfrac{w}{2})}&=&4e^{w}\sum_{l=0}^\infty (l+1) e^{l w} \label{sinh}\ ,\\
\coth(\tfrac{w}{2})&=&-(1+e^{\bar{w}})\sum_{n=0}^\infty  e^{n\bar w}\ .\label{coth}
\eea

Our integral becomes
\bea
I_{C_2}&=&-P^{\dot A \dot B}P^{\dot C\dot D} \epsilon_{\dot C\dot A}\epsilon_{\dot D\dot B}\cr
&&~~\times ~~{1\over 2}\int_{\sigma=0}^{2\pi} d\sigma  \, \left ( {2e^{-{m^2\over 2} w}\over 2^{2m^2}} \right ) \left ( \sum_{k'=0}^{m^2}\, {}^{2m^2}C_{2k'}  e^{k'w}\right ) \left (e^{w}\sum_{l=0}^\infty (l+1) e^{l w}\right )\nn
&&
~~~~~~~~~~~~~~~\times ~~ \left ((1+e^{\bar w})\sum_{n=0}^\infty  e^{n\bar w} \right )\ ,
\label{mone}
\eea
where the last bracket now contains antiholomorphic factors of the form  $1, e^{\bar w}, e^{2\bar w},\cdots$. Following a reasoning similar to that in the case of the computation for $I_{C_1}$ in eq. (\ref{mone_i}), we find that only the leading term, $1$, survives from the last bracket in the limit $T\to \infty$. We then have
\bea
I_{C_2}&=&-P^{\dot C\dot D} P^{\dot A \dot B}\epsilon_{\dot C\dot A}\epsilon_{\dot D\dot B}\,{\pi\over 2^{2m^2}}\sum_{k'=0}^{{m^2\over 2}-1} \, {}^{2m^2}C_{2k'} \, ({\tfrac{m^2}2}-k')\cr
&=&- P^{\dot C\dot D} P^{\dot A \dot B}\epsilon_{\dot C\dot A}\epsilon_{\dot D\dot B}\,{\sqrt{\pi} \over 8}\,{\Gamma[m^2-{1\over2}]\over \Gamma[m^2-1]}=I_{C_1}\ .
\eea

We next consider $I_{C_3}$ in eq. (\ref{IC3}), the contribution from the contour around $w=0$.  For small $|w|$ we have (with $m\ge 2$)
\bea
&&\big(\cosh(\tfrac{w}{4})\big)^{2m^2}+\big(\sinh(\tfrac{w}{4})\big)^{2m^2}=1+{m^2\over 16} w^2+\cdots\ ,\\
&&{1\over \sinh^2(\tfrac{w}{2})}={4\over w^2}-{1\over 3}+\cdots\ ,\label{IC3sinh}\\
&&\coth (\tfrac{w}{2})={2\over \bar w}+{\bar w\over 6}+\cdots\ .\label{IC3cosh}
\eea
The leading term in the integrand (\ref{IC3}) gives:
\be
I_{C_3}\r -P^{\dot A\dot B} P^{\dot C \dot D} \epsilon_{\dot C\dot A}\epsilon_{\dot D\dot B}\,{i\over 2}\int_{|w|=\epsilon} { dw\over w^2}\,{1\over \bar w} = P^{\dot A\dot B}
P^{\dot C \dot D} \epsilon_{\dot C\dot A}\epsilon_{\dot D\dot B}\;{\pi\over \epsilon^2}\ .
\ee
This is a constant independent of the states at the bottom and top of the cylinder  in the correlator, and will arise even if we replace these states with the vacuum state $|0\rangle$. Thus, to maintain the normalization
\be
\langle 0 | 0 \rangle=1
\ee
of the vacuum at $O(\lambda^2)$ we must add to the Lagrangian  a counterterm proportional to the identity operator, which generates an integral
\be
I_{C_3,\,\mathrm{counterterm}}= -P^{\dot A\dot B} P^{\dot C \dot D} \epsilon_{\dot C\dot A}\epsilon_{\dot D\dot B}\,{\pi\over \epsilon^2}\ .
\label{counterterm}
 \ee
  This cancels the divergent contribution to $I_{C_3}$. The next largest terms in the integrand of (\ref{IC3}) give 
\be
P^{\dot A\dot B} P^{\dot C \dot D} \epsilon_{\dot C\dot A}\epsilon_{\dot D\dot B} \left (  -{ im^2\over 32}\int_{|w|=\epsilon} {dw\over \bar w} +  {i\over 24}\int_{|w|=\epsilon} {dw\over \bar w} - {1\over 24}\int_{|w|=\epsilon} {dw  \over  w^2}\,  \bar w  \right )\ .
\ee
We find that each of these terms vanishes. Higher order terms give contributions having positive powers of $|\epsilon|$ and so for these terms we have $I_{C_3,\,\mathrm{renormalized}} = 0$.

Finally, using (\ref{bsix}), we find that for even $m$, $X^{(m)}(T)$ is of the form
\be
\lim_{T\r\infty} X^{(m)}=- P^{\dot A \dot B}P^{\dot C\dot D} \epsilon_{\dot C\dot A}\epsilon_{\dot D\dot B}\,{\sqrt{\pi} \over 4}\,{\Gamma[m^2-{1\over2}]\over \Gamma[m^2-1]}\ .
\ee

\subsection{Computing $X^{(m)}(T)$ for $m$ odd}

We next compute eq. (\ref{bbthree}) for odd values of $m$. Defining $k=2k'+1$, $k'\in\mathbb Z$ we find that
\be
\big(\cosh(\tfrac w4)\big)^{2m^2}-\big(\sinh(\tfrac w4)\big)^{2m^2}={2\over 2^{2m^2}} e^{{(m^2-1)\over 2} w}\sum_{k'=0}^{m^2-1}\, {}^{2m^2}C_{2k'+1}  e^{-k'w}\ .
\ee
Proceeding just as for the case of even $m$, we find
\bea
I_{C_1}&=&- P^{\dot A \dot B}P^{\dot C\dot D}\epsilon_{\dot C\dot A}\epsilon_{\dot D\dot B}\,
{\pi\over 2^{2m^2}}\sum_{k'=0}^{{(m^2-1)\over 2}-1} \, {}^{2m^2}C_{2k'+1} \, ({(m^2-1)\over 2}-k')\cr
&=& - P^{\dot A \dot B}P^{\dot C\dot D} \epsilon_{\dot C\dot A}\epsilon_{\dot D\dot B}\,{\sqrt{\pi} \over 8}\,{\Gamma[m^2-\h]\over \Gamma[m^2-1]}\ .
\eea
We also find, as before,
\be
I_{C_2}=I_{C_1}, ~~~I_{C_3,\,\mathrm{renormalized}}\r 0\ .
\ee
Thus, for odd $m$ we get the same expression as for even $m$
\be
\lim_{T\r\infty} X^{(m)}=- P^{\dot C\dot D} P^{\dot A \dot B}\epsilon_{\dot C\dot A}\epsilon_{\dot D\dot B}\,{\sqrt{\pi} \over 4}\,{\Gamma[m^2-{1\over2}]\over \Gamma[m^2-1]}\ .
\ee

\subsection{Expectation values}

We shall now compute the expectation value of the energy (\ref{deltaE}) for the states (\ref{threex}):
\be
\langle E^{(2)}\rangle=-\pi \lim_{T\r\infty}X^{(m)}= P^{\dot C\dot D} P^{\dot A \dot B}\epsilon_{\dot C\dot A}\epsilon_{\dot D\dot B}\,{\pi^{3\over 2} \over 4}\,
{\Gamma[m^2-{1\over2}]\over \Gamma[m^2-1]}\ .
\ee
We set the polarization $P^{\dot A\dot B}$ to correspond to a perturbation with no quantum numbers
\be
P^{\dot{A}\dot{B}}=\e^{\dot{A}\dot{B}}\ .
\ee
We expect such a perturbation to correspond to the direction towards the supergravity spacetime $AdS_3\times S^3\times T^4$. Then we obtain
\be\label{expval_i}
\langle E^{(2)}\rangle={\pi^{3\over 2} \over 2}\,{\Gamma[m^2-{1\over2}]\over \Gamma[m^2-1]}\ .
\ee
This is the main result of this section. We will generalise this result to arbitrary values of $N$ in section \ref{section7}.

\section{No lift for global modes}\label{section6}

The chiral algebra generators of the CFT at the orbifold point are described by a sum over the generators of each copy:
\be
J^+_{-n}=J^{+(1)}_{-n}+J^{+(2)}_{-n}+\cdots+J^{+(N)}_{-n}\ .
\ee
If we act with such a current on any state then the dimension of the state will rise as 
\be
h\r h+n\ .
\label{azsix}
\ee
There cannot be any anomalous contribution since the change (\ref{azsix}) is determined by the chiral algebra. We can use this fact as a check on the computations that we have performed; we will perform this check for a simple case in this section. 
 
Let us assume that we have  two copies (as in the above sections); so $N=2$.  First consider the case where we apply $J^+_{-1}$; this gives the state
\be
|\chi_1\rangle={1\over \sqrt{2}}J^+_{-1}\, |0\rangle^{(1)}|0\rangle^{(2)}=
{1\over \sqrt{2}}\left ( J^{+(1)}_{-1}+J^{+(2)}_{-1}\right ) \, |0\rangle^{(1)}|0\rangle^{(2)}\ ,
\ee
where we have added a normalization factor to normalize the state to unity. This is the same as the state $|\Phi^{(1)}\rangle$ defined in (\ref{threex}). From (\ref{expval_i}) we see that the lift vanishes in this case.

Next consider the state
\bea
|\chi_2\rangle&=&{1\over \sqrt{8}}J^+_{-3}J^+_{-1} \, |0\rangle^{(1)}|0\rangle^{(2)}\nn
&=&{1\over \sqrt{8}} \left ( J^{+(1)}_{-3}J^{+(1)}_{-1}+J^{+(2)}_{-3}J^{+(2)}_{-1}\right )  \, |0\rangle^{(1)}|0\rangle^{(2)}\cr
&& + {1\over \sqrt{8}}\left ( J^{+(1)}_{-3}J^{+(2)}_{-1}+J^{+(2)}_{-3}J^{+(1)}_{-1}\right )  \, |0\rangle^{(1)}|0\rangle^{(2)}
\nn
&\equiv& |\chi_{2,1}\rangle+|\chi_{2,2}\rangle.
\label{global mode}
\eea
The conjugate state is written in a similar way in two parts
\be
\langle \chi_2|~=~\langle \chi_{2,1}|+\langle \chi_{2,2}|\ .
\ee
Note that the state $|\chi_{2,1}\rangle$ is (upto a normalization factor) the same as the state $|\Phi^{(2)}\rangle$ defined in (\ref{threex}). 

We now consider the lift of the state $|\chi_2\rangle$. There are four contributions to this lift. The first has $|\chi_{2,1}\rangle$ as the initial and final states, and this is proportional to the lift we have computed for $|\Phi^{(2)}\rangle$, see eq. (\ref{expval_i}). Thus, this contribution is nonzero. But there are three other contributions which involve $|\chi_{2,2}\rangle$. When we add all these contributions and subtract the identity contribution in (\ref{axeight}), the lift is expected vanish as we now check. 

Computing the amplitudes by the same method as in the above sections, we find

\bea
 \langle\chi_{2,1}|\big(G^+_{\dot{C},-{1\over2}}\s^-(w_2)\big)\big( G^-_{\dot{A},-{1\over2}}\s^+(w_1)  \big)|\chi_{2,1}\rangle&=&\e_{\dot{C}\dot{A}}{1\over 4}
 \bigg({\cosh^8({\Delta w\over 4}) \over 4\sinh^2({\Delta w\over 2})}  + {\sinh^8({\Delta w\over 4}) \over 4\sinh^2({\Delta w\over 2})}\bigg)\ ,
\cr
\cr
\cr
 \langle\chi_{2,2}|\big(G^+_{\dot{C},-{1\over2}}\s^-(w_2)\big)\big( G^-_{\dot{A},-{1\over2}}\s^+(w_1)  \big)|\chi_{2,1}\rangle  &=& \e_{\dot{C}\dot{A}} {1\over4}
 \bigg( - {\cosh^8 ({\Delta w\over 4})\over 4\sinh^2({\Delta w\over 2})}  - {\sinh^8 ({\Delta w\over 4})\over 4\sinh^2({\Delta w\over 2})}+ \cr
  &&\qquad~~+ {1\over  4\sinh^2({\Delta w\over 2})}\bigg)\ ,
  \cr
  \cr
  \cr
  \langle\chi_{2,1}|\big(G^+_{\dot{C},-{1\over2}}\s^-(w_2)\big)\big( G^-_{\dot{A},-{1\over2}}\s^+(w_1)  \big)|\chi_{2,2}\rangle&=&  \e_{\dot{C}\dot{A}}{1\over4}
  \bigg(-{\cosh^8 ({\Delta w\over 4})\over 4\sinh^2({\Delta w\over 2})} - {\sinh^8 ({\Delta w\over 4})\over 4\sinh^2({\Delta w\over 2})}+\cr
 &&\qquad~~ + {1\over 4\sinh^2({\Delta w\over 2})}\bigg)\ ,
 \cr
 \cr
 \cr
  \langle\chi_{2,2}|\big(G^+_{\dot{C},-{1\over2}}\s^-(w_2)\big)\big( G^-_{\dot{A},-{1\over2}}\s^+(w_1)  \big)|\chi_{2,2}\rangle  &=&\e_{\dot{C}\dot{A}}{1\over4}
  \bigg( {\cosh^8 ({\Delta w\over 4})\over 4\sinh^2({\Delta w\over 2})} +  {\sinh^8 ({\Delta w\over 4})\over 4\sinh^2({\Delta w\over 2})}+\cr
 &&\qquad ~~+  {2\over  4\sinh^2({\Delta w\over 2})}\bigg)\ .
\eea
We add up all four contributions and obtain
\be
 \langle\chi_{2}|\big(G^+_{\dot{C},-{1\over2}}\s^-(w_2)\big)\big( G^-_{\dot{A},-{1\over2}}\s^+(w_1)  \big)|\chi_{2}\rangle=\e_{\dot{C}\dot{A}}{1\over  4\sinh^2({\Delta w\over 2})}\ .
\ee
Collecting the left at right parts of the correlator we find
\be
\langle \chi_2| D(w,\bar{w}) D(0)|\chi_2\rangle = P^{\dot A\dot B} P^{\dot C \dot D} \epsilon_{\dot C\dot A}\epsilon_{\dot D\dot B}
{1\over  16\sinh^2({\Delta w\over 2})}\,{1\over  \sinh^2({\Delta \bar{w}\over 2})}\ .
\ee
Comparing with (\ref{bbtwo}) we have
\be
Q = P^{\dot A\dot B} P^{\dot C \dot D} \epsilon_{\dot C\dot A}\epsilon_{\dot D\dot B}\,{1\over 16}\ .
\label{Q}
\ee
We want to compute $I_{C_1}, I_{C_2}$ and $I_{C_3,\,\mathrm{renormalized}}$ in eqs. (\ref{IC1})-(\ref{IC3}). For $I_{C_1}$, inserting (\ref{Q}) into (\ref{IC1}) yields
\be
I_{C_1}=-P^{\dot A\dot B} P^{\dot C \dot D} \epsilon_{\dot C\dot A}\epsilon_{\dot D\dot B}\int_{\s=0}^{2\pi} d\s \left (  {1\over   16\sinh^2  ({w\over 2})}\,  \coth(\tfrac{\bar w}{2})\right )\ .
\ee
Inserting the expansions (\ref{sinh}) and (\ref{coth}) give
\bea
I_{C_1}=-{1\over4}\int_{\sigma=0}^{2\pi} d\sigma  \,  \left (e^{-w}\sum_{l=0}^\infty {}^{-2}C_l (-1)^l e^{-l w}\right )
\!\!\left ((1+e^{-\bar w})\sum_{n=0}^\infty {}^{-1}C_n(-1)^n e^{-n\bar w} \right )\ . 
\eea

This contour is evaluated at $\tau= {T\over2}$ with $T\to\infty$. Since our expression only contains negative powers of $w$ and $\bar{w}$, every term vanishes and we find that 
\bea
I_{C_1}=0\ .
\eea
Similarly, for $I_{C_2}$ we find
\be
I_{C_2}\,=\,-{1\over4}\int_{\sigma=0}^{2\pi} d\sigma  \, \left (e^{w}\sum_{l=0}^\infty {}^{-2}C_l (-1)^l e^{l w}\right )\!\! \left ((1+e^{\bar w})\sum_{n=0}^\infty {}^{-1}C_n(-1)^n e^{n\bar w}\right)\ .
\ee
This contour is evaluated at $\tau= -{T\over2}$ with $T\to\infty$. Our expression only contains positive powers of $w$ and $\bar{w}$ and we find 
\bea
I_{C_2} = 0\ .
\eea
For $I_{C_3}$, inserting (\ref{Q}) into (\ref{IC3}) yields
\be
 I_{C_3}=-P^{\dot A\dot B} P^{\dot C \dot D} \epsilon_{\dot C\dot A}\epsilon_{\dot D\dot B}i\int_{|w|=\epsilon} dw \left (  {1\over   16\sinh^2  ({w\over 2})}\,  \coth({\bar w\over 2})\right)\ .
\ee
Inserting the expansions (\ref{IC3sinh}) and (\ref{IC3cosh}) and taking the leading order term in the integrand we have
\be
I_{C_3}\r -P^{\dot A\dot B} P^{\dot C \dot D} \epsilon_{\dot C\dot A}\epsilon_{\dot D\dot B}\,{i\over 2}\int_{|w|=\epsilon} { dw\over w^2\bar w}=
P^{\dot A\dot B} P^{\dot C \dot D} \epsilon_{\dot C\dot A}\epsilon_{\dot D\dot B}\,{\pi\over \epsilon^2}\ .
\ee
Using (\ref{counterterm}), we find
\bea
I_{C_{3,\,\mathrm{renormalized}}}=I_{C_3} + I_{C_{3,\,\mathrm{counterterm}}}=0\ .
\eea
The expectation value (\ref{deltaE}) then reads
\bea
\langle E^{(2)}\rangle = -\pi \lim_{T\r\infty}\left ( I_{C_1}+I_{C_2}+I_{C_{3,\,\mathrm{renormalized}}}\right ) = 0\ ,
\eea
proving the absence of lifting for the global mode in (\ref{global mode}). From (\ref{nfour}) we can now argue that the global mode does not mix with any eigenstate that lifts. We have $E^{(2)}_{a'}\ge 0$ for all $a'$, since the states $\t\phi_{a'}$ are chiral primaries in the right-moving sector, and therefore must have $\Delta \bar h =\Delta h \ge 0$. Thus, a vanishing of $\langle E^{(2)}\rangle=0$ means a vanishing overlap of $|\chi_2\rangle$ with each of the $\t\phi_{a'}$ which lift; from this it follows that $|\chi_2\rangle$ remains an unlifted eigenstate of the Hamiltonian.

\section{General values of $N$}\label{section7}

So far we have considered two copies of the $c=6$ CFT: $N=2$. The initial state had two singly-wound copies; the twist operators twisted these together and then untwisted them, so that we ended with two singly-wound copies again. In general, the orbifold CFT has an arbitrary number $N=n_1n_5$ of copies of the CFT. But as we will now see, for our situation, the computation with two copies that we have carried out allows us to obtain the expectation values for arbitrary $N$.

\subsection{The initial state}

We have $N$ copies of the $c=6$ CFT and each copy is singly-wound. Each copy is in the NS sector. Out of these copies, we assume that $n$ copies are excited as
\be
J^{+(i)}_{-(2m-1)}\dots J^{+(i)}_{-3}J^{+(i)}_{-1} |0\rangle^{(i)}\ ,\quad i\in\{1,\cdots,n\}
\ee
in the left moving sector, while the right moving sector is in the vacuum state $|0\rangle$. The remaining $N-n$ copies are in the vacuum state $|0\rangle$ on both the left and right sectors. This state is depicted in fig.\ref{fig_singlwindingexcited}.

There are ${}^NC_n$ ways to choose which strings are excited. Thus, the initial state is composed of ${}^NC_n$ different terms, with each term describing one set of possible excitations. The sum of these terms must be multiplied by a factor
\be
{\cal N}= \left ( {}^NC_n\right ) ^{-\h}
\label{axtone}
\ee
in order that the overall state is normalised to unity.

\subsection{Action of the deformation operator}

When we were dealing with just two copies, we denoted the twist operator of  the deformation by $\sigma^\pm$; it was implicit that this operator would twist together the two copies that we had, see subsection \ref{sebsec_deformation}. But when we have $N>2$ copies, then we need to specify which two copies are being twisted. If the $(i)$ and $(j)$ copies are twisted, we denote the twist operator by $\sigma^\pm_{(i)(j)}$. Thus the deformation operator (\ref{exactlymarginal}) now has the form 
\be
D= P^{\dot A \dot B}G^-_{\dot A, -\h}\bar G^-_{\dot B, -\h} \sum_{i<j}\sigma^{++}_{(i)(j)}\ ,
\ee
where $1\le i,j\le N$. The supercurrents $G^-_{\dot A}$ are given by a sum over the contributions from each copy:
\be
G^-_{\dot A}\equiv\sum_{i=1}^N G^{-(i)}_{\dot A}\ .
\ee

\subsection{Expectation values for general values of $N$}

We argue in the following steps:

\b

(i) Consider the action of the first deformation operator on the initial state. Suppose this first deformation operator twists together the copies $(i),(j)$. Since we are computing an expectation value, the final state must be the same as the initial state; thus the final state must also have all copies singly-wound.  So the second deformation operator must twist the same copies $i,j$ to produce a state with all copies singly-wound. 

\b

(ii) Copies {\it other} than the two copies that get twisted act like `spectators'; thus we get an inner product between their initial state and their final state. If the initial state for such a copy is unexcited, then the final state must also be unexcited, and if the initial state is excited then the final state has to be excited.  The inner product between copies with the same initial and final state is unity.

Since the state of the spectator copies does not change, the number of spectator copies which are excited are also the same between the initial and final states. As a consequence, for the pair $i,j$ which do get twisted, the number of excited copies in the initial and final states is the same. 

\b

(iii) We therefore find that there are three possibilities for the excitations among the twisted copies $i,j$:

\b

(a) Neither of the copies $(i),(j)$ are excited. In this case there is no contribution to the lifting, as the vacuum state $|0\rangle^{(i)}|0\rangle^{(j)}$ is not lifted. 

\b

(b) Both the copies $(i),(j)$ are excited. But the state where both copies are excited is a spectral flow of the state where neither copy is excited, as we have seen in section \ref{section6}. So again there is no lift.

\b

(c) One of the copies out of $(i),(j)$ is excited and one is unexcited. There are four contributions here: (1) Copy $(i)$ excited in the initial state, copy $(i)$ excited in the final state; (2) Copy $(j)$ excited in the initial state, copy $(i)$ excited in the final state; (3) Copy $(i)$ excited in the initial state, copy $(j)$ excited in the final state;  (4) Copy $(j)$ excited in the initial state, copy $(j)$ excited in the final state. These are the four contributions we had in eq. (\ref{adoneqq}). Thus summing these four contributions gives the same anomalous dimension $E^{(2)}$ that we computed for the case $N=2$, see eq. (\ref{expval_i}), with an extra factor of $2$ since we do not here have the normalization factors ${1\over \sqrt{2}}$ in the initial and final state. Thus, the pair $(i),(j)$ contribute $2E^{(2)}$, with $E^{(2)}$ given by eq. (\ref{expval_i}).

\b

(iv) Let us now collect combinatoric factors. First we select the pair $(i),(j)$ out of the $N$ copies. This is done in ${}^NC_2$ ways. Out of these two copies, one has  to be excited (as noted in (iii) above). Thus, out of the remaining $N-2$ copies, $n-1$ are excited. These $n-1$ copies can be chosen in ${}^{N-2}C_{n-1}$ ways, so this is the number of terms which have the contribution $2E^{(2)}$. We note that $1\le n\le N-1$.

\b

(v) Let us finally collect all the factors. We have a normalization factor (\ref{axtone}) both from the initial and final configurations, so we get a factor $|{\cal N}|^2$. Together with the combinatorial factors from the previous paragraph, we find that the expectation value is of the form:
\bea
\langle (E-E_{orbifold})\rangle &=&\lambda^2({}^NC_n) ^{-1}\, ({}^NC_2)\, ({}^{N-2}C_{n-1}) \, (2 \langle E^{(2)}\rangle)\nn
&=&\lambda^2n(N-n)\langle E^{(2)}\rangle \nn
&=&\lambda^2\frac12\,\pi^{3\over 2}\,n(N-n)\,{\Gamma[m^2-{1\over2}]\over \Gamma[m^2-1]}\ .
\eea
The above expression gives the lifting to order $O(\lambda^2)$ for the case where we have $N$ copies of the seed $c=6$ CFT, and $n$ of these are excited by application of the operator ${\cal J}^{(+, m)}$ (eq. (\ref{calj})) which is composed of  $m$ currents. We note that $N=2$ and $n=1$ corresponds to the case studied in section \ref{section5}, see eq. (\ref{expval_i}).

\section{Multi-wound initial states}\label{section8}

So far our initial state has consisted of $N$ singly-wound copies of the seed $c=6$ CFT. We now consider the case where we link together $k$ of these copies to make a `multi-wound copy'. We assume that all copies are grouped into such sets; i.e. there are ${N\over k}$ sets of linked copies, with each set having winding $k$. Note that this requires that $N$ be divisible by $k$. This case is depicted in fig.\ref{fig_multiwound}.

\begin{figure}
\begin{center}
\includegraphics[width=100mm]{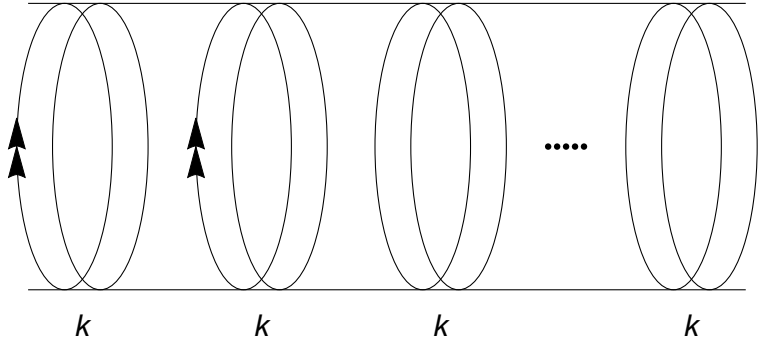}
\end{center}
\caption{ A total of $N$ singly wound copies that have been linked into ${N\over k}$ `multiwound' copies each of winding $k$. A number  $n$ of these twisted sets have been excited by current operators.}
\label{fig_multiwound}
\end{figure}

For the case where the copies are singly-wound, the ground state of each copy was the vacuum $|0\rangle$ with $h=j=0$. A set of linked copies, however, has a  nontrivial dimension, see, e.g. \cite{lm2} for the computation of the ground state energies in both odd and even twisted sectors. We start with each set being in a chiral primary state $|k\rangle$ with
\be\label{cps}
h={k-1\over 2}, ~~~j={k-1\over 2}, ~~~\bar h = {k-1\over 2}, ~~~\bar j = {k-1\over 2}\ .
\ee
We now take $n$ of these linked sets, and excite each of these by the application of current operators:
\be
|k\rangle_{ex}\equiv J^+_{-{(2m-1)\over k}}\dots J^+_{-{3\over k}}J^+_{-{1\over k}}|k\rangle\ .
\label{axttwo}
\ee
This excitation adds a momentum
\be
{1\over k}+{3\over k}+\dots + {(2m-1)\over k}={m^2\over k}
\ee
to this set of linked copies. This momentum must be an integer \cite{dasmathur1}, so $m^2$ should be divisible by $k$.

We wish to find the expectation value of the energy of the state constructed in this way, see eq. (\ref{deltaE}). It turns out that this computation is related to the ones we performed in the above sections by having multiply wound copies in the initial state instead of singly-wound copies. We will now see that we obtain the expectation value for the multi-wound case by going to a covering space of the cylinder, where we undo the multi-winding, and then relating the computation to the singly-wound case.

\subsection{The action of the twist operator}

Consider the action of the first twist operator $\sigma^+_{(i)(j)}$.  There are two possibilities:

\b

(i) The copies $i,j$ are from the same set of linked copies.

\b

(ii) The copies $i,j$ belong to different sets of linked copies.

\b

In case (i), the twist will break up the set of linked copies into two sets with windings $k', k-k'$. Since we are computing an expectation value, the second twist has to link these two sets back to a single set of $k$ linked copies. 

But we can easily see that such an action of twist operators will give no contribution to the expectation value, $\langle E^{(2)}\rangle $. The set of linked copies that we started with could be either unexcited, i.e. in the state $|k\rangle$ (\ref{cps}), or excited, i.e. in the state $|k\rangle_{ex}$ (\ref{axttwo}). The copies other than the ones in our set of $k$ linked copies play no role in the computation. Thus, the action of the two deformation operators tells us the correction $E^{(2)}$ to $|k\rangle$ or $|k\rangle_{ex}$. But $|k\rangle$ is a chiral primary state; this means that its dimension is determined by its charge and so  its anomalous dimension $E^{(2)}$ will have to vanish. The state $|k\rangle_{ex}$ arises from a spectral flow of $|k\rangle$, so again its anomalous dimension will vanish. Thus we get no contributions from the case (i). 

In case (ii), the first twist takes the two sets of $k$ linked copies and  joins them into one set of $2k$ linked copies.  Since we are looking for an expectation value, the second twist must break up this set of $2k$ linked copies back to two sets of copies with linking $k$ each. This is the situation that we will analyze in more detail now.

\subsection{The $k$-fold cover of the cylinder}

Since each set of linked copies has winding number $k$, we can go to a covering space of the cylinder where the spatial coordinate $\t\sigma$ runs over the range $0\le \t \sigma< 2\pi k$. It is convenient to think of the range of $\t\sigma$ to be subdivided into the $k$ intervals
\be
0\le \t\sigma < 2\pi\ , ~~2\pi\le \t\sigma < 4\pi, ~~\dots ~~ 2\pi (k-1)\le \t\sigma < 2\pi k\ .
\ee
We have not changed the time $\tau$ in going to the cover, so we have
\be
\t \tau=\tau\ .
\ee

Now we look at the factors emerging from going to this cover:

\b

(i) For the first set of linked copies, we label the copies $1, 2, \cdots,k$. For the second set, we label them $1', 2', \cdots,k'$. Then the first twist operator has the form $\sigma^+_{(i)(j')}$, where $i$ is from the first set and $j'$ is from the second set. Suppose we consider a twist at the point $\sigma=0$. Then there are  $k^2$ ways of joining the two sets into one set of $2k$ linked copies. 

On the covering space $\t\sigma$, there are $k$ images of the point $\sigma=0$, and we can apply a twist at any of these points. Thus we get $k$ rather than $k^2$ similar interaction points. Thus we must multiply the result we get from the covering space by an extra factor $k$.(The origin of this factor can be understood alternatively as follows: we can choose any of the copies $i=1, \dots k$ to be the first copy $i=1$, and use this to set the origin $\t\sigma=0$; the factor $k$ then describes  the different ways we can choose the  copy $i'=1$ from  the second set of $k$ linked copies.)

\b

(ii) The second twist acts on a set of $2k$ linked copies labeled by $i=1, \cdots,2k$. Suppose this twist is at the location $\sigma=0$ on the cylinder. This time there are only $k$ possible ways for the twist to act: once we choose one of the copies $i$, the second copy must be $i'=i+k$, since otherwise the set will not break up into two sets of winding $k$ each. Because the twist is symmetric  between $i$ and $i'$, the different possibilities are given by choosing $i=1, \dots k$; i.e., there are $k$ possible choices. 

We now see that on the covering space $\t\sigma$, these $k$ different choices are accounted for by the $k$ different images of $\sigma=0$ on the space $\t\sigma$. Thus there is no additional factor (analogous to case (i)) from the second twist. 

\b

(iii) We now make a conformal map  from the covering space $\t\sigma, \t\tau$ to a cylinder where the spatial coordinate has the usual range $(0, 2\pi)$:
\be
\sigma'={1\over k}\,{\t\sigma}\ , ~~~~\tau'={1\over k}\,\t\tau\ .
\ee
Under this map the deformation operators scale as
\be
D(\t w_1, \bar {\t w}_1) \r  {1\over k^2} D(w'_1, \bar {w'}_1)\ , ~~~D(\t w_2, \bar {\t w}_2) \r  {1\over k^2} D(w'_2, \bar {w'}_2)\ ,
\ee
so we get an overal factor of ${1\over k^4}$ from this scaling.


\b

(iv) Let us now recall the computation of  the amplitude $A^{(2)}(T)$ in subsection \ref{subsec_Aab}, see eq.s (\ref{wnasel_ii}) and (\ref{A2T_i}). This amplitude involved integrals over the positions $w_1, w_2$  of the two deformation operators. (We later recast these integrals in the form of contour integrals in eq. (\ref{A2T_i}), but it is simpler to see the scalings in terms of the original integrals in eq. (\ref{wnasel_ii})). We have
\be
\int d^2 \t w_1 \r k^2 \int  d^2 w'_1\ ,\qquad\int d^2 \t w_2 \r k^2 \int  d^2 w'_2\ ,
\ee
so we obtain a overall factor of $k^4$.

\b

(v) The integral over $\Delta \t w=(\t w_2-\t w_1)$ converges, but the integral over $s=\h(\t w_1+\t w_2)$ gives a factor of $\t T$. We need to multiply by a factor ${1\over \t T}$, which scales as
\be
{1\over \t T} \r {1\over k}\,{1\over T'}\ .
\label{axtfiveqq}
\ee

\b

(vi) We have now mapped the problem to the cylinder $w'$, which is just like the cylinder $w$ which worked with in the situation with $k=1$.  Collecting all the factors we obtained from (i)-(v) above, we find that the factors of $k$ cancel out. We thus find the following: suppose we have $N=n_1n_5$ copies of the CFT. These copies are grouped into ${N\over k}$ sets of copies, with each set having $k$ linked copies. A number $0\le n\le {N\over k}$ of these sets is excited in the form $|k\rangle_{ex}$, while the remainder are in the state $|k\rangle$. Then the lifting of the energy is given by
\be
\langle (E-E_{orbifold})\rangle=\lambda^2\frac{\pi^{3\over 2}}2\,n(N-n)\,{\Gamma[m^2-{1\over2}]\over \Gamma[m^2-1]}\ .
\ee
The above expression gives the lifting to order $O(\lambda^2)$ for the case where we have $N$ copies of the $c=6$ CFT, with these copies being grouped into ${N\over k}$ sets (with $N$ being divisible by $k$) with each set  having winding $k$. Of these sets,  $n$ are excited in the form (\ref{axttwo}) which describes the action of  $m$ fractionally moded currents.

\section{The maximally wound sector}\label{section9}

We have computed the lifting of certain states which are excited on the left, but are a chiral primary on the right. Apart from special cases this lifting was found to be nonzero. 
In \cite{gavanarain}, the lifting of a more general class of  states was computed in a certain approximation; again it was found that generic states were lifted. 

Let us analyze this lifting in the context of the elliptic genus \cite{Witten:1993jg} which tells us how many unlifted states we expect at a given energy for the left movers. The elliptic genus for the case where the compactification was $K3\times S^1$ was computed in \cite{deBoer:1998kjm,deBoer:1998us}. The elliptic genus vanishes for the compactification $T^4\times S^1$ that we have considered, but a modified index was defined in \cite{Maldacena:1999bp}. This index protects very few states for low levels of the left moving energy. Thus in the Ramond (R) sector, there are very few unlifted states for 
\be
h\le {N\over 4}\ .
\ee
But for $h>{N\over 4}$ the number of states that are unlifted is very large; in fact their number ${\cal N}$ has to reproduce the black hole entropy which behaves as
\be
S=\ln {\cal N} \approx 2\pi \sqrt{Nh}\ .
\label{azone}
\ee
Thus we need to ask: what changes when we cross the threshold $h={N\over 4}$? Since we have been working in the NS sector, let us first spectral flow to the NS sector. Consider the R sector ground state with maximal twist $k=N$. This state has dimension
\be
h={c\over 24}={N\over 4}\ .
\ee
and charge $j=\pm \h$; let us take $j=-\h$. The spectral flow of this state to the NS sector gives a  chiral primary with
\be
h=j={N-1\over 2}\ .
\ee
This is the state $|k\rangle$ we defined above with $k=N$. We can obtain states contributing to the entropy (\ref{azone}) by acting with left moving creation and annihilation operators on $|N\rangle$. 

 Let us now ask if there is a special property shared by states in  the maximally wound sector, which is not present for states in sectors where we do not have maximal winding. We will now argue that there is indeed such a property: the nature of the linkage that is produced by the action of twists.
 
 If a state is not in the maximally wound sector, then the twist $\sigma_{(i)(j)}$ present in the deformation operator can do one of two things: 
 
 \b
 
 (i) It can join two different set of linked copies, with windings $k_1, k_2$,  into one linked copy with winding $k_1+k_2$. The second deformation operator will break this back to two sets with windings $k_1, k_2$, since we are computing an expectation value and so need the final state to be the same as the initial state.
 
 \b
 
 (ii) If we have a subset of strings with winding $k_1+k_2>1$, then it can break this subset into two sets of linked copies, with windings $k_1, k_2$. The second deformation operator will join these sets back to one set with winding $k_1+k_2$. 
 
 \b
 
 If on the other hand we have a state in the maximally wound sector, then there are no other sets of copies in the state; thus we are allowed possibility (ii) but not possibility (i).
 
  We must now ask if there is a difference in the action of the deformation operators in the cases (i) and (ii). We will see that there is indeed a difference:  in case (i), the covering space obtained when we `undo' the twist operators is a sphere (genus $g=0$), while in case (ii) the covering space is a torus (genus $g=1$). 
  
  To see this, we recall how we compute the genus of the covering space obtained from undoing the action of twist operators \cite{lm1}. Suppose we have twists of order $k_i, i=1, \dots i_{max}$. The ramification order at a twist $\sigma_{k_i}$ is $r_i=k_i-1$. Let the number of sheets (i.e. copies) over a generic point be ${ s}$. Then the genus of the covering surface is given by the Riemann-Hurwitz relation
  \be
  g=\h \sum_i r_i -s+1\ .
  \ee
  
  Let us now compute $g$ in the two cases above. We focus only on the copies which are involved in the interaction:
  
  \b
  
  (i') In case (i), we create the initial set of linked strings using twist operators $\sigma_{k_1}, \sigma_{k_2}$. The final state is created by the twists of the same order. The two deformation operators carry twists $\sigma_2$ each. The number of sheets is $s=k_1+k_2$. Thus
  \be
  g=\h [ 2(k_1-1)+2(k_2-1)+2]-(k_1+k_2)+1=0\ .
  \ee

  \b
  
  (ii') In case (ii), we create the initial state by a twist $\sigma_{k_1+k_2}$.  The final state is created by a twist of the same order. The two deformation operators carry twists $\sigma_2$ each. The number of sheets is $s=k_1+k_2$. Thus
  \be
  g=\h [ 2(k_1+k_2-1)+2 ]-(k_1+k_2)+1=1\ .
  \ee

In the present paper we have considered an example of case (i), where the covering space $t$ was a sphere. In this situation we found that the lift $\langle E^{(2)}\rangle$ was nonzero. It is possible that when the covering space is a torus, then the lift vanishes, at least for some class of states. If that happens, then such states in the maximally wound sector will not be lifted. We hope to return to this issue elsewhere.

\section{Discussion}\label{section10}

We have considered the family of states depicted in fig.\ref{fig_singlwindingexcited}, and computed the correction to the expectation value of their  energy - the `lift' - upto second order in the deformation parameter $\lambda$. The results, depicted in fig.\ref{fig_deltaE}, suggest a heuristic picture for this lift; this picture was discussed in section \ref{introsec}. 

We know that the lift vanishes in two extreme cases: (i) when no copies are excited and (ii) when all copies are excited. (The state in (ii) is just a spectral flow of the state in (i).) But in between these two extremes, the energy does rise. The heuristic picture aims to explain this phenomenon as follows.

Each set of linked copies corresponds, in this heuristic picture, to one elementary object in the dual gravity configuration. The excitations (\ref{threex}) for  $m>1$ do not correspond to supergravity quanta; thus we must think of them as `string' states. 
String states will have more  mass  than charge, and will therefore `lift'. But the gravitational attraction between the strings will cause the overall energy to reduce.   If we have enough strings so that all the copies in the CFT are excited, then the negative potential energy cancels the energy from string tension, and we end up with no lift.

We noted in section \ref{section8} that this picture holds also for the case where the copies of the $c=6$ CFT are linked together in sets of $k$ copies each. Thus if none of these sets is excited then we have no lift, and again if all the sets are excited we have no lift. But when some of the sets are excited, then we do have a lift in general. 

Now consider the limiting case where {\it all} the $N$ copies of the CFT are linked into one copy with winding $k=N$. If we excite this multi-wound set, then we have excited {\it all} the sets, since there is only one set to excite. If we extrapolate our heuristic picture above to this limiting case, then we see that the energy lift of this string state will be cancelled by the self-gravitation of the state, and the state will not lift at all. This suggests that states in the maximally wound sector will not lift.\footnote{It has been argued earlier \cite{maldasuss} that states relevant for the dynamics of the {\it near-extremal} hole should be in the highly wound sectors.} This is interesting, because we know that at high energies we have to reproduce the large entropy of the extremal hole \cite{sv}, so we need a large class of states that will not lift.

This picture also tells us how we should think about states in the fuzzball paradigm. There are certainly some states in general winding sectors that are not lifted, and the gravity description of many of these states have been constructed. But as we have seen, many states {\it will} lift. The fuzzball paradigm says that {\it all} states are fuzzballs; i.e., they have no regular horizon. Thus the class of states obtained in the fuzzball construction will in general cover both extremal and non-extremal states. It turns out that it is often easier to take a limit where the non-extremal states are in fact near-extremal. We can then look for the subclass of extremal states as limits of the construction that gives the near-extremal states. 

We also note that one should not make a sharp distinction between `supergarvity' states and
`stringy' states. In fig.\ref{fig_deltaE} the state with $n=0$ is a supergravity state with no strings. As $n$ rises, we add more an more strings, but at $n=N$ this collection of strings again behaves like a supergravity state with no lift.

We have suggested above that a large class of extremal states might lie in the maximally wound sector. We noted in section \ref{section8}  that the $O(\lambda^2)$ lift of such states  has contributions only from  genus $1$ covering surfaces while states with lower winding have both genus $0$ and genus $1$ contributions; this fact may be relevant to the relation between lifting and maximal winding. But maximally wound sectors are difficult to study in the classical limit: the classical limit corresponds to  $N\r\infty$, and a winding $k=N$ will typically produce a conical defect with conical angle $1/k=1/N\r 0$ \cite{bal,mm,glmt}. Thus such states should be thought of as limits of states with finite $k$. If the general $k$ states are near-extremal, and the $k=N$ state is extremal, then the extremal state can be seen as a limit of a family of near-extremal states.

Finally, we note that we have considered only a special family of D1-D5-P states in this paper.  It is of interest to ask if there is a general characterization of which D1-D5-P states lift and which do not. We hope to study this issue elsewhere.

\section*{Acknowledgements}
We would like to thank Shouvik Datta, Lorenz Eberhardt, Matthias Gaberdiel, Christoph Keller, Alessandro Sfondrini, and David Turton for helpful discussions. We especially thank Stefano Giusto and Rodolfo Russo for extended discussions on this problem. IGZ thanks the STAG Research Centre at Southampton University for hospitality and the organisers of the \emph{Workshop on holography, gauge theories and black holes} for the stimulating environment. The work of SH and SDM is supported in part by the DOE grant DE-SC0011726. The work of IGZ is supported by the Swiss National Science Foundation through the NCCR SwissMAP.

\begin{appendix}
\section{Notation and conventions} \label{app_cft}
\subsection{Field Definitions}
Here we give the notation and conventions used in our computations.
We have 4 real left moving fermions $\psi_1, \psi_2, \psi_3, \psi_4$ which are groupped into doublets $\psi^{\alpha A}$ as follows:
\be
\begin{pmatrix}
\psi^{++} \cr \psi^{-+}
\end{pmatrix}
={1\over\sqrt{2}}
\begin{pmatrix}
\psi_1+i\psi_2 \cr \psi_3+i\psi_4
\end{pmatrix}
\ee
\be
\begin{pmatrix}
\psi^{+-} \cr \psi^{--}
\end{pmatrix}
={1\over\sqrt{2}}
\begin{pmatrix}
\psi_3-i\psi_4 \cr -(\psi_1-i\psi_2)
\end{pmatrix}.
\ee
The index $\alpha=(+,-)$ corresponds to the subgroup $SU(2)_L$ of rotations on $S^3$ and the index $A=(+,-)$ corresponds to the subgroup $SU(2)_1$ from rotations in $T^4$. The 2-point functions read
\be
\langle\psi^{\alpha A}(z)\psi^{\beta B}(w)\rangle=-\epsilon^{\alpha\beta}\epsilon^{AB}{1\over z-w}
\ee
where we have 
\be
\epsilon_{12}=1, ~~~\epsilon^{12}=-1
\ee
The 4 real left-moving bosons $X_1, X_2, X_3, X_4$ are grouped into a matrix 
\be
X_{A\dot A}= {1\over\sqrt{2}} X_i \sigma_i
={1\over\sqrt{2}}
\begin{pmatrix}
X_3+iX_4 & X_1-iX_2 \cr X_1+iX_2&-X_3+iX_4
\end{pmatrix}
\ee
where $\sigma_i=(\sigma_a, iI)$. The bosonic field 2-point functions are then of the form
\be
\langle\pa X_{A\dot A}(z) \pa X_{B\dot B}(w)\rangle=\epsilon_{AB}\epsilon_{\dot A\dot B}{1\over (z-w)^2} \,.
\ee
The chiral algebra is generated by the R-currents, supercurrents, and the stress-energy tensor:
\bea
J^a&=& {1\over 4}\e_{\a \g}\e_{AC}\psi^{\g C} (\sigma^{Ta})^\alpha{}_\beta \psi^{\beta A},\qquad a=1,2,3
\cr\cr
G^\alpha_{\dot A}&=& \psi^{\alpha A} \pa X_{A\dot A},\qquad \a = +,-
\cr\cr
T&=& {1\over 2} \e^{AB}\e^{\dot{A}\dot{B}}\pa X_{B\dot B}\pa X_{A\dot A} + {1\over 2} \e_{\a\beta}\e_{AB}\psi^{\beta B} \pa \psi^{\alpha A}
\eea

\subsection{OPE Algebra}
We note the OPEs between the various operators of interest.

\subsubsection{OPE's of currents with $\partial X_{A\dot{A}}(z)$ and $\psi^{\a A}(z)$ }

\bea
T(z)\partial X_{A\dot{A}}(w) &\sim& {\partial X_{A \dot A}(w)\over (z-w)^2} + { \partial^2 X_{A \dot A}(w)\over z-w} \cr
T(z)\psi^{\a A}(w) &\sim&  {{1\over 2}\psi^{\a A}(w)\over (z-w)^2}+{\partial \psi^{\a A}(w)\over z-w} \cr
G^{\a}_{\dot{A}}(z)\psi^{\beta B}(w) &\sim& \e^{\a\beta}\e^{BA} {\partial X_{A\dot{A}}(w) \over z-w}\cr
G^{\a}_{\dot{A}}(z)\partial X_{ B\dot{B}}(w)&\sim& \e_{AB}\e_{\dot{A}\dot{B}}{ \psi^{\a A}(w)\over  (z-w)^2} + \e_{AB}\e_{\dot{A}\dot{B}}{ \partial \psi^{\a A}(w)\over  z-w}\cr
J^a(z)\psi^{\a A}(w) &\sim&  {1\over 2} {1\over z-w}(\s^{Ta})^{\a}_{\beta}\psi^{\beta A}(w)\cr
J^{+}(z)\psi^{+ A}(w) &=& 0,\qquad J^{-}(z)\psi^{+ A}(w) =   {\psi^{- A}(w)\over z-w} \cr
J^{+}(z)\psi^{- A}(w) &=&   {\psi^{+ A}(w)\over z-w} ,\qquad  J^{-}(z)\psi^{- A}(w) = 0
\eea

\subsubsection{OPE's of currents with currents }

\bea
T(z)T(w)&\sim&{{c\over2}\over (z-w)^4} + {2T(w)\over (z-w)^2}  + {\partial T(w)\over z-w}
\cr
\cr
J^a(z)J^b(w)&\sim&  {{c\over 12}\delta^{ab}\over(z-w)^2} + {i\e^{ab}_{\,\,\,\,c}J^c(w)\over z-w}
\cr
\cr
G^{\a}_{\dot{A}}(z)G^{\beta}_{\dot{B}}(w) &\sim& -\e_{\dot{A}\dot{B}}\bigg[\e^{\beta\a}{{c\over3}\over (z-w)^3}  + \e^{\beta\g}(\s^{aT})^{\a}_{\g}\bigg({2J^a(w)\over (z-w)^2} + {\partial J^a(w)\over z-w}\bigg)  + \e^{\beta\a}{1\over z-w}T(w)   \bigg]
\cr
\cr
J^a(z)G^{\a}_{\dot{A}}(w)&\sim& {1\over z-w}{1\over2}(\s^{aT})^{\a}_{\beta}G^{\beta}_{\dot{A}}(w)
\cr
\cr
T(z)J^a(w) &\sim&  {J^a(w)\over (z-w)^2} + {\partial J^a(w)\over z-w}
\cr
\cr
T(z)G^{\a}_{\dot{A}}(w)&\sim& {{3\over2}G^{\a}_{\dot{A}}(w)\over (z-w)^2} + { \partial G^{\a}_{\dot{A}}(w)\over z-w}
\eea
We convert the relations involving $J^1, J^2$ to those involving $J^+, J^-$. Defining $J^{+}, J^-$ as
\bea
J^+ &=& J^1 + i J^2\cr
J^-&=& J^1 - i J^2
\eea
yield the following OPE's
\ba
J^{+}(z)J^{-}(w)&\sim  {{c\over6}\over (z-w)^2} +  {2J^3(w)\over z-w},& J^{-}(z)J^{+}(w)&\sim  {{c\over6}\over (z-w)^2} -  {2J^3(w)\over z-w}\cr
J^3(z)J^{+}(w)&\sim  {J^{+}(w)\over z-w}, & J^3(z)J^{-}(w)&\sim- {J^{-}(w)\over z-w}\cr
J^+(z)J^3(w)&\sim- {J^+(w)\over z-w}, &J^-(z)J^3(w)&\sim {J^-(w)\over z-w}\cr
T(z)J^{+}(w)&\sim{J^{+}(w)\over (z-w)^2} + {\partial J^{+}(w)\over z-w},&T(z)J^{-}(w)&\sim {J^{-}(w)\over (z-w)^2} + {\partial J^{-}(w)\over z-w}\cr
J^{+}(z)G^{-}_{\dot{A}}(w) &\sim {G^{+}_{\dot{A}}(w)\over z-w},& J^{-}(z)G^{+}_{\dot{A}}(w)& \sim {G^{-}_{\dot{A}}(w)\over z-w}
\end{align}

\subsection{Mode and contour definitions of the fields} 

The modes are defined in terms of contours through
\bea
L_m&=&\oint {dz\over 2\pi i}z^{m+1}T(z)\cr
J^a_m&=&\oint {dz\over 2\pi i} z^{m}J^a(z)\cr
G^{\a}_{\dot{A},r}&=&\oint {dz\over 2\pi i} z^{r+{1\over2}}G^{\a}_{\dot{A}}(z)\cr
\a_{A\dot{A},m}&=&i\oint {dz\over 2\pi i} z^{m}\partial X_{A\dot{A}}(z)\cr
d^{\a A}_r&=& \oint {dz\over 2\pi i} z^{r-{1\over2}}\psi^{\a A}(z)
\eea 

The inverse relations are
\bea
T(z) &=& \sum_{m}z^{-m-2}L_m\cr
J^a(z) &=& \sum_{m}z^{-m-1}J^a_m\cr
G^{\a}_{\dot{A}}(z) &=& \sum_{r}z^{-r-{3\over2}}G^{\a}_{\dot{A},r}\cr
\partial X_{A\dot{A}}(z) &=& -i\sum_m z^{-m-1}\a_{A\dot{A},m}\cr
\psi^{\a A}(z) &=& \sum_m z^{-m-{1\over2}}d^ {\a A}_m
\eea

\subsection{Commutation relations}
\subsubsection{Commutators of $\alpha_{A\dot{A},m}$ and $d^{\a A}_r$}

\bea
[\a_{A\dot{A},m},\a_{B\dot{B},n}] &=& -m\e_{A\dot{A}}\e_{B\dot{B}}\delta_{m+n,0}\cr
[d^{\alpha A}_r , d^{\beta B}_s]  &=&-\e^{\alpha\beta}\e^{AB}\delta_{r+s,0}
\eea

\subsubsection{Commutators of currents with $\alpha_{A\dot{A},m}$ and $d^{\a A}_r$}
\bea\label{commutations}
[L_m,\a_{A\dot{A},n}] &=&-n\a_{A\dot{A},m+n} \cr
[L_m ,d^{\a A}_r] &=&-({m\over2}+r)d^{\a A}_{m+r}\cr
\lbrace G^{\a}_{\dot{A},r} ,  d^{\beta B}_{s} \rbrace&=&i\e^{\a\beta}\e^{AB}\a_{A\dot{A},r+s}\cr
[G^{\a}_{\dot{A},r} , \a_{B \dot{B},m}]&=&  -im\e_{AB}\e_{\dot{A}\dot{B}}d^{\a A}_{r+m}\cr
[J^a_m,d^{\a A}_r] &=&{1\over 2}(\s^{Ta})^{\a}_{\beta}d^{\beta A}_{m+r}\cr
[J^{+}_m,d^{+ A}_r] &=& 0,\qquad~~~~~ [J^{-}_m,d^{+ A}_r] ~=~ d^{-A}_{m+r}\cr
[J^{-}_m,d^{+ A}_r] &=& d^{-A}_{m+r},\qquad [J^{+}_m,d^{+ A}_r] ~=~ 0
\eea

\subsubsection{Commutators of currents with currents}
\bea\label{commutations_ii}
[L_m,L_n] &=& {c\over12}m(m^2-1)\delta_{m+n,0}+ (m-n)L_{m+n}\cr
[J^a_{m},J^b_{n}] &=&{c\over12}m\delta^{ab}\delta_{m+n,0} +  i\e^{ab}_{\,\,\,\,c}J^c_{m+n}\cr
\lbrace G^{\a}_{\dot{A},r} , G^{\beta}_{\dot{B},s} \rbrace&=&  \e_{\dot{A}\dot{B}}\bigg[\e^{\a\beta}{c\over6}(r^2-{1\over4})\delta_{r+s,0}  + (\s^{aT})^{\a}_{\g}\e^{\g\beta}(r-s)J^a_{r+s}  + \e^{\a\beta}L_{r+s}  \bigg]\cr
[J^a_{m},G^{\a}_{\dot{A},r}] &=&{1\over2}(\s^{aT})^{\a}_{\beta} G^{\beta}_{\dot{A},m+r}\cr
[L_{m},J^a_n]&=& -nJ^a_{m+n}\cr
[L_{m},G^{\a}_{\dot{A},r}] &=& ({m\over2}  -r)G^{\a}_{\dot{A},m+r}\cr
[J^+_{m},J^-_{n}]&=&{c\over6}m\delta_{m+n,0} + 2J^3_{m+n}\cr
[L_m,J^{+}_n] &=& -nJ^{+}_{m+n},\qquad ~[L_m,J^{-}_n] ~=~ -nJ^{-}_{m+n}\cr
[J^{+}_{m},G^{+}_{\dot{A},r}]  &=& 0 ,\qquad\qquad ~~~[J^{-}_{m},G^{+}_{\dot{A},r}]  ~=~ G^{-}_{\dot{A},m+r}\cr
[J^{+}_{m},G^{-}_{\dot{A},r}]  &=&G^{+}_{\dot{A},m+r},\qquad ~[J^{-}_{m},G^{-}_{\dot{A},r}]  ~=~ 0 \cr
[J^3_m , J^{+}_n] &=& J^{+}_{m+n},\qquad\qquad [J^3_m , J^{-}_n] ~=~ -J^{-}_{m+n}
\eea

\subsection{Current modes written in terms of $\alpha_{A\dot{A},m}$ and $d^{\a A}_r$}
\bea
J^a_m &=& {1\over 4}\sum_{r}\epsilon_{AB}d^ {\g B}_r\epsilon_{\alpha\gamma}(\s^{aT})^{\a}_{\beta}d^ {\beta A}_{m-r},\qquad a=1,2,3\cr
J^3_m &=&  - {1\over 2}\sum_{r} d^ {+ +}_{r}d^ {- -}_{m-r} - {1\over 2}\sum_{r}d^ {- +}_r d^ {+ -}_{m-r}\cr
J^{+}_m&=&\sum_{r}d^ {+ +}_rd^ {+ -}_{m-r} ,\qquad J^{-}_m=\sum_{r}d^ {--}_rd^ {- +}_{m-r}\cr
G^{\a}_{\dot{A},r} &=& -i\sum_{n}d^ {\a A}_{r-n} \a_{A\dot{A},n}\cr
L_m&=& -{1\over 2}\sum_{n} \e^{AB}\e^{\dot A \dot B}\a_{A\dot{A},n}\a_{B\dot{B},m-n}- {1\over 2}\sum_{r}(m-r+{1\over2})\epsilon_{\alpha\beta}\epsilon_{AB}d^ {\a A}_r d^ {\beta B}_{m-r}
\eea

\section{Spectral Flow}\label{app_sf}
In this appendix we review the rules for spectral flow transformations \cite{Schwimmer:1986mf}. Under spectral flow by $\a$ units, the dimension, $h$, and the charge, $j$, transform like

\bea
h' = h +\alpha j +{c\a^2\over 24}\ , \qquad j' = j +{\a c\over12}\ ,
\eea
where $c$ is the central charge of the CFT. Consider an operator $\mathcal{O}(z)$ of charge $q$. Under spectral flow by $\a$ units at a point $z_0$, the operator transforms as
\bea\label{sf_O}
\mathcal{O}(z)\to (z-z_0)^{-\a q}O(z)\ .
\eea

\end{appendix}

\end{document}